\documentclass[11pt,reqno,preprint]{article}
\usepackage{jheppub}
\usepackage{amssymb,amsmath}
\usepackage{color}
\usepackage{hyperref}
\usepackage{graphicx}
\usepackage{bbold}
\usepackage{MnSymbol}
\usepackage{subcaption}
\usepackage{caption}
\title{Bit models of replica wormholes}
\author{Henry Maxfield}
\affiliation{Stanford Institute for Theoretical Physics, Stanford University, Stanford, CA 94305, USA}
\emailAdd{henrym@stanford.edu}

\abstract{We define bit models of evaporating black holes which incorporate the effects of replica wormholes. These enter as non-perturbative corrections to the gravitational inner product, arising from spacetime wormholes that connect replica black holes. The resulting models have exactly unitary evolution, predict measurements on Hawking radiation in accord with a Page curve for entropy, and are compatible with a smooth horizon for infalling observers.}

\newcommand{\hilb}{\mathcal{H}}	
\newcommand{\hbu}{\mathcal{H}_\text{BU}} 
\newcommand{\CC}{\mathbb{C}}		
\newcommand{\id}{\mathbb{1}}

\usepackage{mathrsfs}

\newcommand{\HH}{\mathrm{HH}} 

\newcommand{\dmanifold}{\mathcal{M}} 

\newcommand{\scri}{\mathscr{I}}
\newcommand{\op}{\mathcal{O}}
\newcommand{\island}{\mathcal{I}}
\newcommand{\constr}{\mathbf{\eta}}

\DeclareMathOperator{\Tr}{Tr}

\DeclareMathOperator{\Sym}{Sym}

\graphicspath{{./Figures/}}

\begin{document}
\maketitle

\section{Introduction and summary}

Black holes present a theoretical challenge to the foundations of quantum gravity, since two facts are in tension. The first is that semi-classical gravity appears to admit an unbounded number of states in the interior of black holes of any given mass. The second is that the microscopic interpretation of Bekenstein-Hawking entropy (supported in particular by the AdS/CFT correspondence and examples of state counting in string theory \cite{Strominger:1996sh,Dabholkar:2014ema}) bounds the same number of states. A quantitative version of this tension is the Page curve \cite{Page:2013dx}, a prediction for the entropy of Hawking radiation as a function of time. The process of Hawking radiation generates entanglement between interior states and the radiation leading to an ever-growing entropy, which eventually comes into conflict with the bound from the Bekenstein-Hawking entropy, given by the horizon area in Planck units $S_\mathrm{BH}= \frac{A}{4G_N}$.

It is often useful to clarify a problem by reducing it to the simplest possible model while still retaining essential features, and the black hole information problem is no exception. A useful class of models for black holes was introduced by Mathur \cite{Mathur:2009hf} and further explored by \cite{Giddings:2011ks,Giddings:2012dh,Polchinski:2016hrw}.  These model the interior state of a black hole by a sequence of quantum bits, and the process of Hawking radiation creates a new bit (increasing the volume of the black hole interior) in a maximally entangled state with a corresponding bit of Hawking radiation. Such models illustrate the essential point that small corrections to this entangled state of Hawking radiation can never give rise to the decreasing entropy of radiation required by the Page curve, sometimes called the `small corrections theorem' \cite{Mathur:2009hf}. Large corrections are interpreted as a violation of semi-classical physics, in particular a failure to have a smooth horizon. This idea was further sharpened by the firewall argument \cite{Almheiri:2012rt,Almheiri:2013hfa}.

However, it has recently been understood that gravity does in fact predict the Page curve once particular non-perturbative effects are taken into account \cite{Penington:2019npb,Almheiri:2019psf,Penington:2019kki,Almheiri:2019qdq}. This result requires fluctuations in the topology of spacetime, in the form of `replica wormhole' geometries which connect the interiors of several identical replica black holes entering into the calculation of entropy. The resulting calculations are in precise, quantitative agreement with the expectation that black holes behave like ordinary quantum systems with Hilbert space dimension $e^{S_\mathrm{BH}}$ (sometimes called the `central dogma' \cite{Almheiri:2020cfm}).

Nonetheless, the replica wormhole computations of the entropy do not directly tell us much about the Hilbert space of black hole states: can we describe the states in geometric gravitational language, and why is the large Hilbert space of perturbative quantum gravity incorrect? Answers to these questions seem necessary to discuss the dynamics inside and near black holes, the required modifications to the state of Hawking radiation, or the experience of infalling observers.

In this paper we address some of these questions in the simplified context of a bit model, modifying Mathur's by incorporating replica wormholes. The resulting model concretely describes an exactly unitary time evolution on the black hole and its radiation which simultaneously achieves a smooth horizon and a Page curve: any bit of Hawking radiation is in a maximally entangled state with its interior partner up to exponentially small corrections, while all measurements on the Hawking radiation are consistent with an entropy bounded by $S_\mathrm{BH}$.

The key idea is to consider replica wormhole contributions not only to entropy calculations, but to a more basic quantity: the inner product on the black hole Hilbert space. We do this in the most straightforward and standard way, by cutting open a path integral on a spatial slice and defining a Hilbert space by wavefunctions of field configurations on that slice.\footnote{This is more subtle with gravity than on a fixed spacetime background, since fixing a spatial slice involves a choice of gauge under diffeomorphisms, and gauge invariance imposes the gravitational constraints. This does not raise any conceptual difficultly in perturbation theory around a given background, but non-perturbative consequences will be crucial!} In particular, we still interpret states as wavefunctions of the usual gravitational variables, the metric and matter fields; the only modification is to the inner product. This has the virtue of more direct contact with operators in the black hole interior (for example, there is no need for a dual CFT and discussion of bulk reconstruction relevant to that context).

 However, the result of following the standard procedure to define a Hilbert space from the path integral has an unusual feature due to wormholes: it violates cluster decomposition (the property that inner products on composite systems factorise in the limit of large separation between the components), though only in a very mild way that does not endanger validity of ordinary local physics. This connects to the ideas of baby universes, superselection sectors (called $\alpha$-states), and (from an AdS/CFT point of view) duality with an ensemble \cite{Giddings:1988cx,Coleman:1988cy,Saad:2019lba,Marolf:2020xie}. Nonetheless, even if one is sceptical of such a scenario (understandably, given top-down examples of AdS/CFT), we believe that the main idea should be taken seriously. Namely, non-perturbative corrections to inner products computed in a standard manner from low-energy gravitational effective field theory can have profound implications for the Hilbert space of quantum gravity.

\subsection{Non-perturbative contributions to the inner product}\label{sec:NPIP}

Before diving into the concrete models in the main part of the paper, we set the stage with a sketch of the main idea.

Our aim is to compute an inner product in a theory of gravity between states defined on some Cauchy surface $\Sigma$, here phrased in the language of the path integral. In general, to compute the inner product between two states we perform the path integral over all geometries bounded by two copies of $\Sigma$, on which the two states in question define boundary conditions for the fields (including metric). In a classical limit (in particular, with small curvatures in Planck units) we typically expect this to be dominated by perturbations around some fixed background, with Hilbert space $\hilb_\mathrm{pert}$ of the states $|\psi\rangle$ of quantum fields on $\Sigma$. These fields should include small metric fluctuations, so in particular $|\psi\rangle$ will respect the gravitational constraints order-by-order in perturbation theory around the given background (and some gauge choice uniquely identifies the slice $\Sigma$ for small perturbations around the background). We imagine that we have thus defined an inner product $\langle\phi|\psi\rangle$ order by order in perturbation theory.

But this perturbative construction   $\hilb_\mathrm{pert}$ may not give a complete picture of the Hilbert space: the  inner product might receive non-perturbative corrections, giving us a physical Hilbert space $\hilb_\mathrm{phys}$ with modified inner product $\llangle\phi|\psi\rrangle$. The corrections come from summing over additional geometries $\dmanifold$  that are not small perturbations of the original background. 

The contribution of any given $\dmanifold$ to the inner product comes from two places. First, the classical gravitational action $S_\mathrm{grav}[\dmanifold]$ contributes a factor $\lambda_\dmanifold =e^{iS_\mathrm{grav}[\dmanifold]} $; the manifolds we consider will turn out to have an imaginary Lorentzian action, meaning that $\lambda_\dmanifold$ is exponentially small in $G_N^{-1}$ and so $\lambda_\dmanifold\ll 1$ typically suppresses the new contributions. Secondly, we have the path integral of perturbative quantum fields on the new background $\dmanifold$ with boundary conditions determined by the states $|\psi\rangle$, $|\phi\rangle$. This matter amplitude can be represented as a matrix element $\langle \phi|\eta_\dmanifold|\psi\rangle$ of some operator $\eta_\dmanifold$ acting on the perturbative Hilbert space $\hilb_\mathrm{pert}$. Combining these, we can write the physical inner product as
\begin{equation}
\begin{aligned}
	\llangle \phi|\psi\rrangle &= \sum_\dmanifold \lambda_\dmanifold \langle \phi|\eta_\dmanifold|\psi\rangle \\
	&=\langle \phi|\eta|\psi\rangle  \qquad \left(\eta = \sum_\dmanifold \lambda_\dmanifold \eta_\dmanifold \right), \label{eq:IPconstr1}
\end{aligned}	
\end{equation}
where the sum\footnote{This sum over $\dmanifold$ might include a continuous integral over some moduli (associated with zero modes or nearly-zero modes for the metric perturbations) while the perturbative inner product $\langle \phi|\eta_\dmanifold|\psi\rangle$ takes care of fluctuations orthogonal to such moduli.} runs over all relevant geometries $\dmanifold$.  Of course, without a complete microscopic definition of the gravitational path integral we can never be sure exactly what class of geometries $\dmanifold$ should be included:  our aim is only to suggest a particular class, namely replica wormholes, and explore the consequences.

To implement this idea in bit models, we intepret $\hilb_\mathrm{pert}$ as the Hilbert space of a collection $\hilb_k$ of  $k$ qbits essentially as in the original model \cite{Mathur:2009hf}, with the slight modification that we will take $n$ copies (or `replicas') of this Hilbert space so $\hilb_\mathrm{pert}=\hilb_k^{\otimes n}$. We then interpret replica wormholes as the spacetimes which compute matrix elements of permutation operators $\eta_\mathcal{M} = \hat{\pi}(\island)$: these operators act as permutations $\pi\in\Sym(n)$ of the $n$ replicas, acting on some subset $\island$ of the bits. $\dmanifold$ labels $\pi$ and $\island$. Finally, we are free to choose the coefficients $\eta_\dmanifold$, though we motivate their form by considering the gravitational action of replica wormholes.

Such contributions will be sufficient to simultaneously achieve three aims. First, time evolution of the black hole (which produces a bit of Hawking radiation in an entangled state with a new interior mode) is implemented by an exactly unitary operator. Secondly, all measurements on Hawking radiation will be compatible with a Page curve; in particular, a decreasing entropy as more bits of radiation are collected. Finally, measurements on the interior modes (such as verifying the correct entangled state between radiation and interior partners corresponding to a smooth horizon) receive exponentially small corrections compared the perturbative results.


\subsection{Outline}

The remainder of the paper is organised as follows.

The main work is in \textbf{\autoref{sec:model}}, where we motivate and define the bit models. The staring point is essentially the same as the model of \cite{Mathur:2009hf}, described in section \ref{sec:singleBH}. Section \ref{sec:RWIP} discusses the path integral on Lorentzian replica wormhole geometries, which motivates the definition of the replica inner product in bit models in section \ref{sec:RWmodels}. Some key properties of the resulting models (in particular their unitary evolution) are described in section \ref{sec:properties}. The bit models have free parameters, so we conclude the section by describing some concrete choices: a very simple version in section \ref{sec:PS} that we call the `Polchinski-Strominger model' (following ideas of \cite{Polchinski:1994zs}), and a family motivated by comparison with more realistic models of replica wormholes in section \ref{sec:RWmodel}.

In order to discuss the information problem in the bit models, we must address two key points: restoring information in the radiation (via the Page curve), and semiclassical physics in the interior. We tackle these in turn, beginning with the Page curve in \textbf{\autoref{sec:Page}}. We introduce the operationally-defined notion of entropy relevant for measurements of asymptotic observers, and show that it follows the Page curve (via a version of the `quantum extremal surface formula') in our models.

We then turn to physics in the interior of the black hole in section \textbf{\autoref{sec:smooth}}. States in our model are expressed in terms of  the usual  semiclassical gravitational variables, so one might expect that it is straightforward to discuss interior observables. However, modifying the inner product does introduce some subtleties. These arise because non-perturbative effects lead to ambiguities in the definition of bulk operators \cite{Jafferis:2017tiu}. We give one natural construction of non-perturbative interior operators, and show that it achieves the aim of small corrections of matrix elements relative to the perturbative values in an appropriate semiclassical regime.

Next in \textbf{\autoref{sec:BUs}} we address the unusual feature of our inner products alluded to above: namely that they do not factorise between separate black holes. We show that a weaker notion of factorisation is satisfied; namely that the Hilbert space splits as a direct sum of superselection sectors in which factorisation does hold. With this property, failure of factorisation does not threaten usual local physics for measurements outside the horizon. We explain why this property is guaranteed on general grounds via an inner product on a Hilbert space of `baby universes'.

Finally, \textbf{\autoref{sec:disc}} gives an extensive discussion of various ideas and open questions which arise from the model, broken up into subsections.

\section{The model}\label{sec:model}

\subsection{A bit model of a single black hole}\label{sec:singleBH}

We first describe the model for a single black hole. By definition, replica wormholes involve multiple copies (the replicas), so will not play any part in this description of a single system. This section will therefore describe a conventional model of a semiclassical evaporating black hole, very similar to (indeed, inspired by) \cite{Mathur:2009hf}.

\begin{figure}
\centering
\begin{subfigure}[b]{.45\textwidth}
	\includegraphics[width=.9\textwidth]{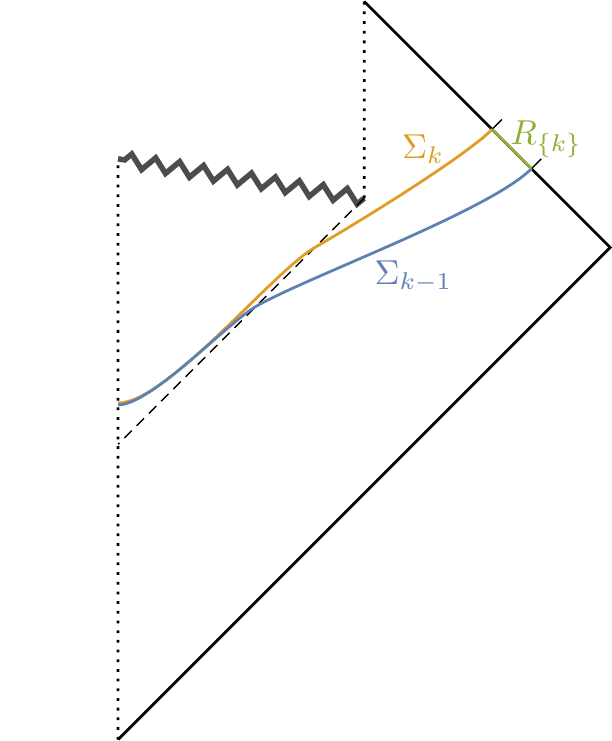}
	\caption{Carter-Penrose}
\end{subfigure}
\begin{subfigure}[b]{.45\textwidth}
\centering
	\includegraphics[width=.7\textwidth]{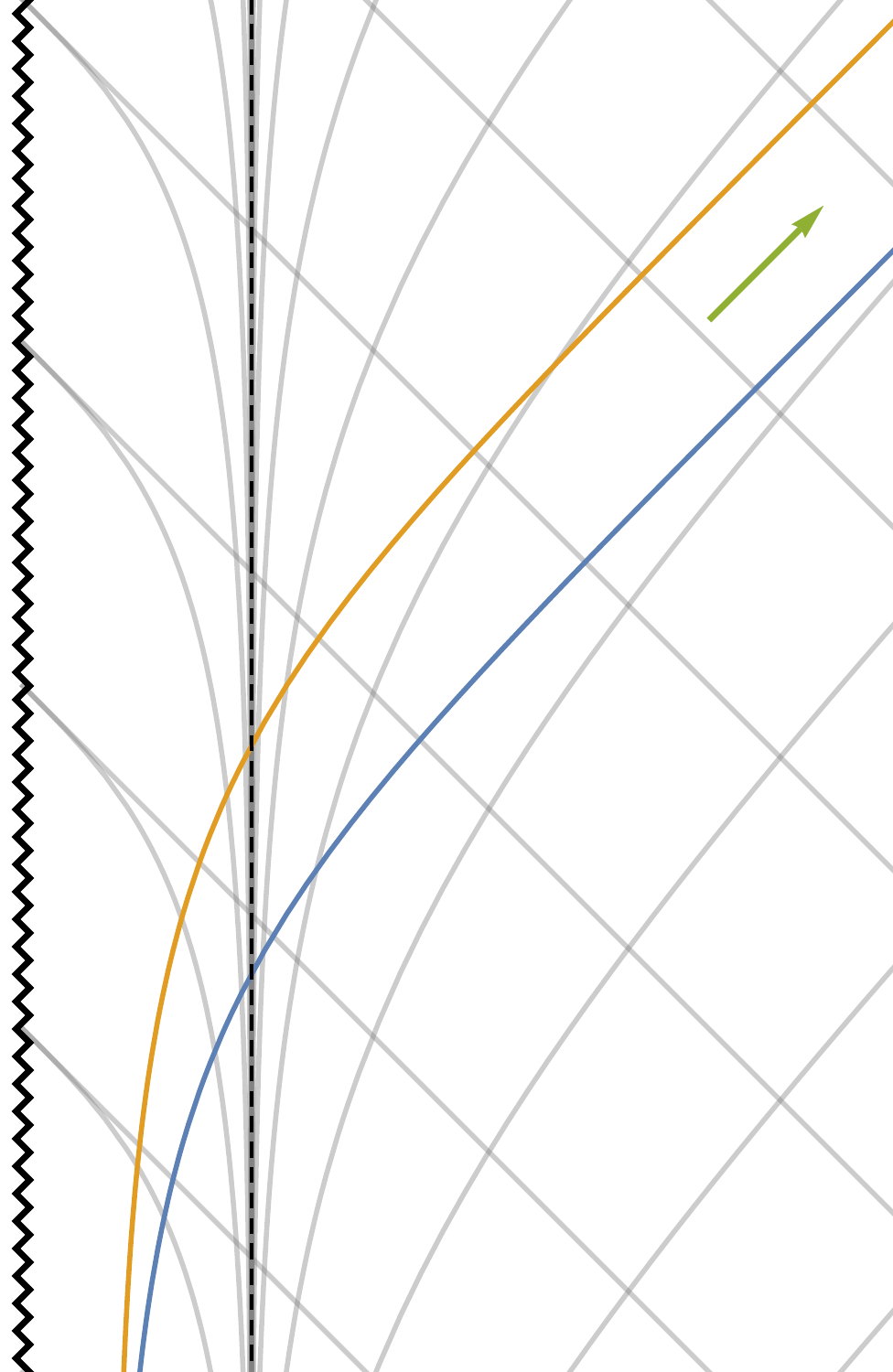}
	\caption{Eddington-Finkelstein}
\end{subfigure}
	\caption{The Hilbert spaces $\hilb_k$ of our bit models represent states on the `nice slices' $\Sigma_k$. On the left two such slices are shown on a Penrose diagram (so ingoing and outgoing null rays move at 45 degrees). On the right we show the same using ingoing Eddington-Finkelstein coordinates (ingoing null rays move at 45 degrees, but outgoing do not; the null geodesics are shown in gray). This makes the approximate time-translation symmetry manifest as vertical translations. The map $U_k$ takes states on $\Sigma_{k-1}$ to states on $\Sigma_k$ along with a bit of Hawking radiation $R_{\{k\}}$, which is shown in green. The jagged line indicates the curvature singularity. \label{fig:niceSlice}}
\end{figure}

The first piece of the model consists of a sequence of Hilbert spaces $\hilb_k$ indexed by a non-negative integer $k$, which labels a discrete sequence of times $t_k$. A state in $\hilb_k$ models the state of a black hole formed from collapse on a `nice slice' $\Sigma_k$ as depicted in figure \ref{fig:niceSlice}.
The slice $\Sigma_k$ is a partial Cauchy surface covering the interior of the black hole up to some time $t_k$
, but chosen so that it does not contain the Hawking radiation emitted before that time. So, the states in $\hilb_k$ model the state of matter that forms the black hole along with the state of `Hawking partners', outmoving excitations in the interior created by the Hawking process (which we describe in a moment). We do not include in $\hilb_k$ the state of any emitted Hawking radiation; this will be contained in a separate Hilbert space factor. We do not explicitly model any of the ingoing modes such as the matter that collapsed to form the black hole in the first place (though it is simple to generalise to include this): we may take these modes to always be in some fixed pure state.
 
Beyond this the details are not so important, but we offer a more concrete description for further orientation, with spherically symmetric black holes in mind. We may take part of $\Sigma_k$ for retarded times $v\lesssim t_k$ to lie inside the black hole at approximately constant radial coordinate $r$ (less than the horizon radius). To include replica wormholes later, it will be convenient to take this surface to lie just inside the event horizon, with area less than that of the event horizon by an amount of order $G_N$ (specifically, we can take this to be the locus of quantum extremal surfaces), though this is not essential.
For larger retarded times $v\gtrsim t_k$, we take $\Sigma_k$ to be approximately null, lying close to the surface of advanced time $u\approx t_k$, and terminating at $\scri^+$ (or at the timelike asymptotic boundary in AdS). These two pieces are connected smoothly in the region near the horizon (the `zone') by a surface crossing the horizon at retarded time $v\approx t_k$. See \cite{Mathur:2009hf} for a more detailed description of similar slices (which differ in that the exterior portion lies at constant Schwarzschild time, instead of nearly null as here). If there is no matter falling into the black hole at times greater than $t_k$ (as we will assume throughout), the state on the portion of the slice outside the horizon is close to the vacuum, so the state on $\Sigma_k$ is indeed a description of the state of the interior only.

Over time, the volume of the slices $\Sigma_k$ behind the horizon steadily grows, and the number of possible low-energy excitations grows commensurately.  The discrete label $k$ counts the number of such interior modes, so that the state on $\Sigma_k$ requires $k$ bits to describe. One may think of $k$ as a proxy for the volume of the black hole interior, or as a discrete time variable. This is related to the physical time steps $t_k-t_{k-1}$ by a factor of order the inverse temperature $\beta$ (though $t_k$ and $k$ will not be precisely proportional over long times, since $\beta$ will vary as the black hole shrinks).

\begin{figure}
\centering
\begin{subfigure}[b]{.45\textwidth}
	\includegraphics[width=.9\textwidth]{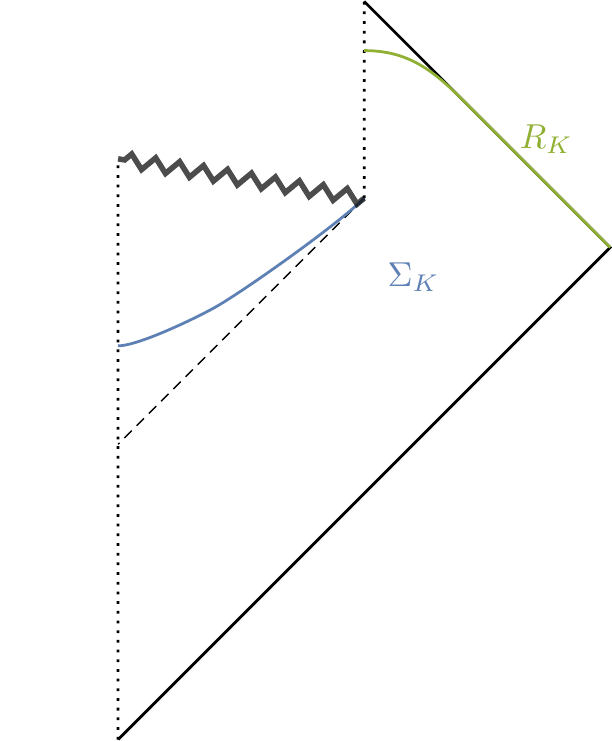}
	\caption{Penrose diagram}
\end{subfigure}
\begin{subfigure}[b]{.5\textwidth}
\centering
	\includegraphics[width=\textwidth]{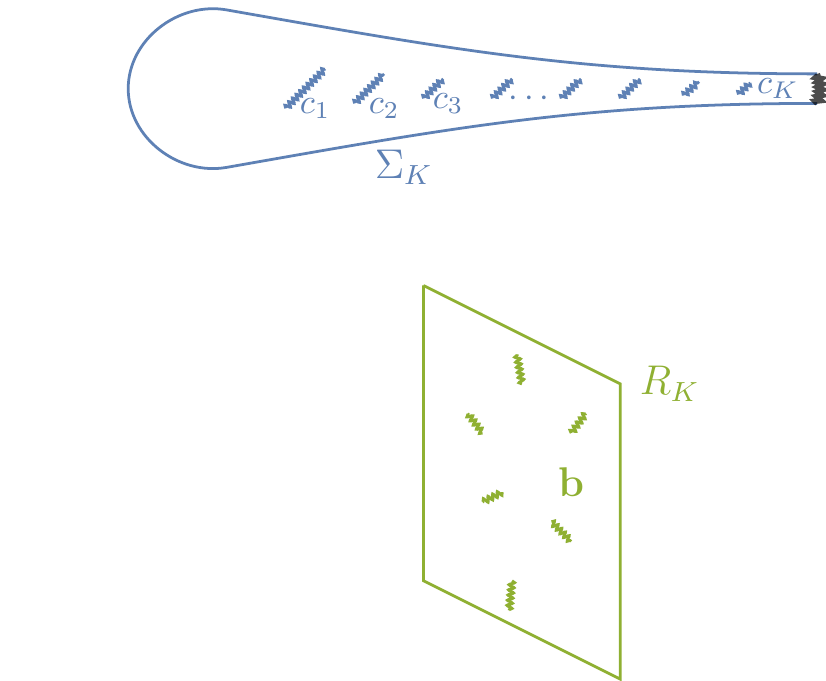}
    \caption{Spatial geometry}
\end{subfigure}
	\caption{We obtain a convenient simplification by assuming that the black hole evaporates completely after emitting $K$ bits of Hawking radiation. The post-evaporation state is defined on a final `Cauchy surface' which splits into two disconnected components, pictured here on a Penrose diagram (left) and as a cartoon of the spatial geometry (right). One component is an almost-flat Minkowski space exterior $R_K$, containing all the outgoing Hawking radiation $\mathbf{b}$ modes. The other component is the black hole interior $\Sigma_K$, containing the infalling state and Hawking partner  $\mathbf{c}$ modes.  $\Sigma_K$ can be interpreted as a `closed universe' (a compact Cauchy surface with no asymptotic boundary). The geometry on  $\Sigma_K$ is weakly curved except for a Planck-sized region associated with the endpoint of evaporation (indicated by the jagged lines on the right figure), though evolution will shortly end at the black hole singularity everywhere. The endpoint is unimportant except for the assumptions that the black hole evaporates completely, and that the dynamics in the final moments of evaporation are local (in particular, independent of the initial size of the black hole).
	\label{fig:finalSlice}}
\end{figure}

The black hole will eventually evaporate completely after emitting $K \gg 1$ bits of radiation, giving us a maximal value of the index $k$: $0\leq k\leq K$. In particular,  we may think of the final slice $\Sigma_K$ as a partial Cauchy surface covering the black hole interior as pictured in \ref{fig:finalSlice} (we may ignore the null piece on the exterior since the ingoing state is vacuum by assumption). This final surface does not play an important role in general, but is useful to interpret the simplest version of the replica wormhole model introduced in section \ref{sec:PS}.

This this motivation in mind, we define the Hilbert space $\hilb_k$ (modelling states on $\Sigma_k$) as a collection $\mathbf{c}$ of $k$ bits $c_1,c_2,\ldots,c_k$:
\begin{gather}
	\hilb_k = \operatorname{span} \big\{ |\mathbf{c}\rangle = |c_1,\ldots,c_k\rangle, c_i \in \{\uparrow ,\downarrow\} \big\},  \\
	\langle \mathbf{c}'|\mathbf{c}\rangle = \delta_{\mathbf{c}'\mathbf{c}} := \delta_{c_1'c_1} \cdots \, \delta_{c'_kc_k}  \label{eq:HkIP}\,.
\end{gather}
 One can easily generalise this by taking the labels $c_i$ to run over any number of values, but we stick to the simplest case of bits for concreteness. The Hilbert space for the first slice $\hilb_0$ is one-dimensional, spanned by a single state $|\:\rangle$ representing some definite pure state of the ingoing modes (the collapsing star that forms the initial black hole, say).  We can imagine creating any possible states in $\hilb_k$ by allowing the black hole to evaporate, performing measurements on the radiation and post-selecting on the results.

The second piece of the model is a time evolution that takes us from a state in $\hilb_{k-1}$ to a state in a joint system consisting of $\hilb_{k}$ along with a factor modelling the the Hawking radiation emitted in that time. The state on $\Sigma_{k-1}$ will not evolve unitarily to the state on $\Sigma_k$ alone, since these two slices end at different asymptotic times. Instead, it evolves to a state on $\Sigma_k \cup R_{\{k\}}$, where $R_{\{k\}}$ is the portion of $\scri^+$ between times $t_{k-1}$ and $t_k$. The Hilbert space $\hilb_{R_{\{k\}}}$ for $R_{\{k\}}$ contains the Hawking radiation emitted in that interval of time, which we take to be encoded in a single bit --- the $k$th bit of radiation --- so $\hilb_{R_{\{k\}}}$ is two-dimensional. The state on $\Sigma_k \cup R_{\{k\}}$ consists of the original state on $\Sigma_{k-1}$ (unaltered), along with a new interior excitation and the emitted bit of Hawking radiation, which we take to be maximally entangled. The evolution is therefore described by an operator $U_k$, defined by
\begin{gather}
	U_k:\hilb_{k-1}\to \hilb_k\otimes \hilb_{R_{\{k\}}}\\
	U_k|\mathbf{c}\rangle = \sum_{b_k,c_k} \psi_{b_k c_k}|\mathbf{c},c_k\rangle\otimes |b_k\rangle, \qquad \sum_{c_k}\psi_{b_k c_k}\psi_{b_k' c_k}^* = \tfrac{1}{2}\delta_{b_k b_k'}. \label{eq:U}
\end{gather}
The coefficients $\psi_{b_kc_k}$ describe the wavefunction of the Hawking radiation and interior partner; the last equality says $\psi_{b_kc_k}$ is a maximally entangled state on two bits. We could write $\psi_{b_kc_k} = \frac{1}{\sqrt{2}}\delta_{b_kc_k}$ by choice of basis, but we leave it arbitrary for clarity in distinguishing between interior and exterior modes. This generalises straightforwardly to take $b_k,c_k$ to run over any number of labels and  for any state $\psi_{b_kc_k}$; our specific evolution is chosen for concreteness and simplicity.

We may use repeated applications of $U_{k+1}$, $U_{k+2}$, \dots $U_l$ to evolve from any slice $\Sigma_k$ to any later slice $\Sigma_l$ ($l> k$), emitting $l-k$ bits of radiation into a Hilbert space $\hilb_{R_{\{k+1,k+2,\ldots,l-1,l\}}} =\hilb_{R_{\{k+1\}}}\otimes \cdots\otimes \hilb_{R_{\{l\}}} $. For the first $k$ bits of radiation we use the shorthand $\hilb_{R_k} = \hilb_{R_{\{1,2,\ldots,k\}}}$, capturing the radiation emitted when evolving from $\Sigma_0$ to $\Sigma_k$. We will also simply write $U$ for the evolution operator between consecutive slices, with the index $k$ implied by the space on which it acts. In particular, $U^k$ acts as $U_kU_{k-1}\cdots U_1$ on $\hilb_0$, taking the initial state on $\Sigma_0$ to a state on $\Sigma_k \cup R_k$:
\begin{equation}
\begin{aligned}
	|\psi_k\rangle = U^k |\phantom{c}\rangle &=\sum_{b_i,c_i}\psi_{b_1c_1}\cdots \psi_{b_kc_k} |c_1c_2\cdots c_k\rangle_{\hilb_k} \otimes |b_1b_2\cdots b_k\rangle_{R_k}  \\
	&=\sum_{\mathbf{b},\mathbf{c}}\psi_{\mathbf{b}\mathbf{c}} |\mathbf{c}\rangle_{\hilb_k} \otimes |\mathbf{b}\rangle_{R_k},
\end{aligned}
\end{equation}
where the second line defines the useful shorthand notation $\psi_{\mathbf{b}\mathbf{c}} = \psi_{b_1c_1}\cdots \psi_{b_kc_k}$ for the full wavefunction of $k$ bits of Hawking radiation. If we trace out the black hole, we find that the radiation density matrix $\rho_{k}=\rho_{R_k}$ is a maximally mixed state on the $2^k$-dimensional radiation Hilbert space $\hilb_{R_k}$:
\begin{equation}
\begin{aligned}
	\rho_{k} &= \Tr_{\Sigma_k}\left(|\psi_k\rangle\langle\psi_k|\right) \\
	&= \sum_{\mathbf{b},\mathbf{c},\mathbf{b}',\mathbf{c}'}  \bar{\psi}_{\mathbf{b}'\mathbf{c}'}\psi_{\mathbf{b}\mathbf{c}} \langle \mathbf{c'}|\mathbf{c} \rangle  \; |\mathbf{b}\rangle\langle\mathbf{b}'| \\
	&= \frac{1}{2^k} \id.
\end{aligned}
\end{equation}
The second line emphasises that the state of radiation depends on the inner product on the black hole Hilbert space, defined in \eqref{eq:HkIP} as $\langle \mathbf{c'}|\mathbf{c} \rangle = \delta_{\mathbf{c'}\mathbf{c}}$. This inner product is the only part of the model that we will modify when we incorporate replica wormholes.

Note that $U_k$ as defined here is not strictly a unitary operator, but only an isometry: $U_k^\dag U_k = \id$, where $\id$ is the identity on $\hilb_{k-1}$, so $U_k$ preserves inner products, but $U_k U_k^\dag $ does not equal the identity on $\hilb_k\otimes \hilb_{R_{\{k\}}}$, instead projecting onto the maximally entangled state of $b_k,c_k$. This occurs because $\hilb_{k-1}$ does not attempt to describe all possible states on $\Sigma_k$, but only those states without any high-energy excitations. This property is preserved by forward time evolution but not backward evolution, since backward evolution blueshifts generic low-energy excitations (as might result from disrupting the entanglement between $b_k$ and $c_k$) near the horizon. So, we should bear in mind that $U_k$ attempts to model only `forward' evolution, and not the reverse evolution of a general state backwards to an earlier slice.

\subsection{Motivation: replica wormhole contributions to the inner product}\label{sec:RWIP}

Now we would like to modify the bit model to incorporate replica wormholes. Specifically, we want to implement the idea of section \ref{sec:NPIP}, allowing the path integral over such spacetimes to contribute modifications to the inner product on the Hilbert space of multiple black holes. To motivate this, we discuss the geometry of replica wormhole Lorenztian spacetimes, mostly follow the ideas of \cite{Marolf:2020rpm} (summarised in \cite{Marolf:2021ghr}) to which we refer for a more detailed exposition.


One way to define a replica wormhole $\dmanifold$ is via the operator $\eta_\dmanifold$ computed by the path integral of perturbative quantum fields on $\dmanifold$. First, we should describe the space of states $\hilb_\mathrm{pert}$ on which $\eta_\dmanifold$ acts. These are states on $n$ `replica' black holes, each defined on an identical Cauchy surface $\Sigma_k$ as in the discussion above, so $\Sigma$ is the disjoint union of the $n$ copies of $\Sigma_k$. These states live in the $n$-fold tensor product Hilbert space $\hilb_\mathrm{pert} = \hilb_k^{\otimes n}$.

Now, our replica manifolds $\dmanifold$ are labelled by two things: one is a partial Cauchy surface $\island$ in the single replica spacetime, referred to as the `island', and the other is a permutation $\pi\in \Sym(n)$ of the $n$ replicas. The path integral on $\dmanifold(\island,\pi)$ then computes the operator $\hat{\pi}(\island)$ which acts by permuting fields on the island:
\begin{equation}
	\eta_\dmanifold = \hat{\pi}(\island).
\end{equation}
We unpack the details of this operator in the following.

A partial Cauchy surface is a region of space: a spacelike achronal compact hypersurface, with boundary $\partial\island$ (called the splitting surface) which is co-dimension two in spacetime. Alternatively, since $\hat{\pi}(\island)$ depends only on the maximal causal development of $\island$ we can label our operator by a causal diamond.   The splitting surface $\partial\island$ divides a complete Cauchy surface into a union $\island \cup \bar{\island}$ of the island and its complement sharing the common boundary $\partial\island$. If the Hilbert space $\hilb_k$ of quantum fields on each replica splits as a tensor product $\hilb_\island \otimes \hilb_{\bar{\island}}$ spanned by states $|\psi_r\rangle \otimes |\bar{\psi}_r\rangle$ (where $r=1,\ldots,n$ labels the replica), we can write the action of $\hat{\pi}(\island)$ very explicitly:
 \begin{equation}\label{eq:permop}
 \begin{gathered}
 		\hat{\pi}(\island) \text{ acts on }\hilb_\mathrm{pert} = \hilb_\island \otimes \hilb_{\bar{\island}}\otimes\cdots\otimes \hilb_\island \otimes \hilb_{\bar{\island}}  \text{ as}\\
 		\hat{\pi}(\island) |\psi_1\rangle \otimes |\bar{\psi}_1\rangle\otimes \cdots \otimes |\psi_n\rangle \otimes |\bar{\psi}_n\rangle = |\psi_{\pi^{-1}(1)}\rangle \otimes |\bar{\psi}_1\rangle\otimes \cdots \otimes |\psi_{\pi^{-1}(n)}\rangle \otimes |\bar{\psi}_n\rangle.
 \end{gathered}
 \end{equation}
 
These operators may sound unfamiliar, but they are in fact closely related to standard computations of entropies in QFT using the replica trick \cite{Holzhey:1994we,Headrick:2019eth} . Specifically, the expectation value of $\hat{\pi}(\island)$ on the $n$-fold tensor product of some state $\rho$ computes the $n$th R\'enyi entropy $S_n(\island)$ of the island in the state $\rho$:
\begin{equation}\label{eq:pientropy}
	\Tr(\rho^{\otimes n} \hat{\pi}(\island)) = \Tr(\rho_\island^n) = e^{-(n-1)S_n(\island)},
\end{equation}
where $\pi\in\Sym(n)$ is a cyclic permutation (such as $\pi=\tau_n$, defined as $ \tau_n(1)=n$, $\tau_n(r)=r-1$ for $r=2,\ldots n$). Here, $\rho_\island$ is the reduced density matrix for fields on the island, obtained from the partial trace over its complement: $\rho_\island = \Tr_{\bar{\island}}\rho$. A `replica wormhole' is just the geometry upon which we perform the  replica trick path integral to evaluate this quantity.
 
 Importantly, the island $\island$ need not be a region contained within the original Cauchy slice $\Sigma_l$. For us, it may be a region in another Cauchy  slice $\Sigma_{k_\island}$ which lies to the future of $\Sigma_k$ (so $k_\island>k$). In that case,  $\hat{\pi}(\island)$ should be interpreted as a Heisenberg operator acting on $\hilb_k^{\otimes n}$: it acts simply (as in \eqref{eq:permop}) on $\hilb_{k_\island}^{\otimes n}$, but to obtain its action on $\hilb_k^{\otimes n}$ one must conjugate by ($n$ copies of) the time-evolution operator $U:\hilb_k\to \hilb_{k_\island}$ which relates the Hilbert spaces defined on these different Cauchy slices.
 
 \begin{figure}[!h]
\centering
	\includegraphics[width=.7\textwidth]{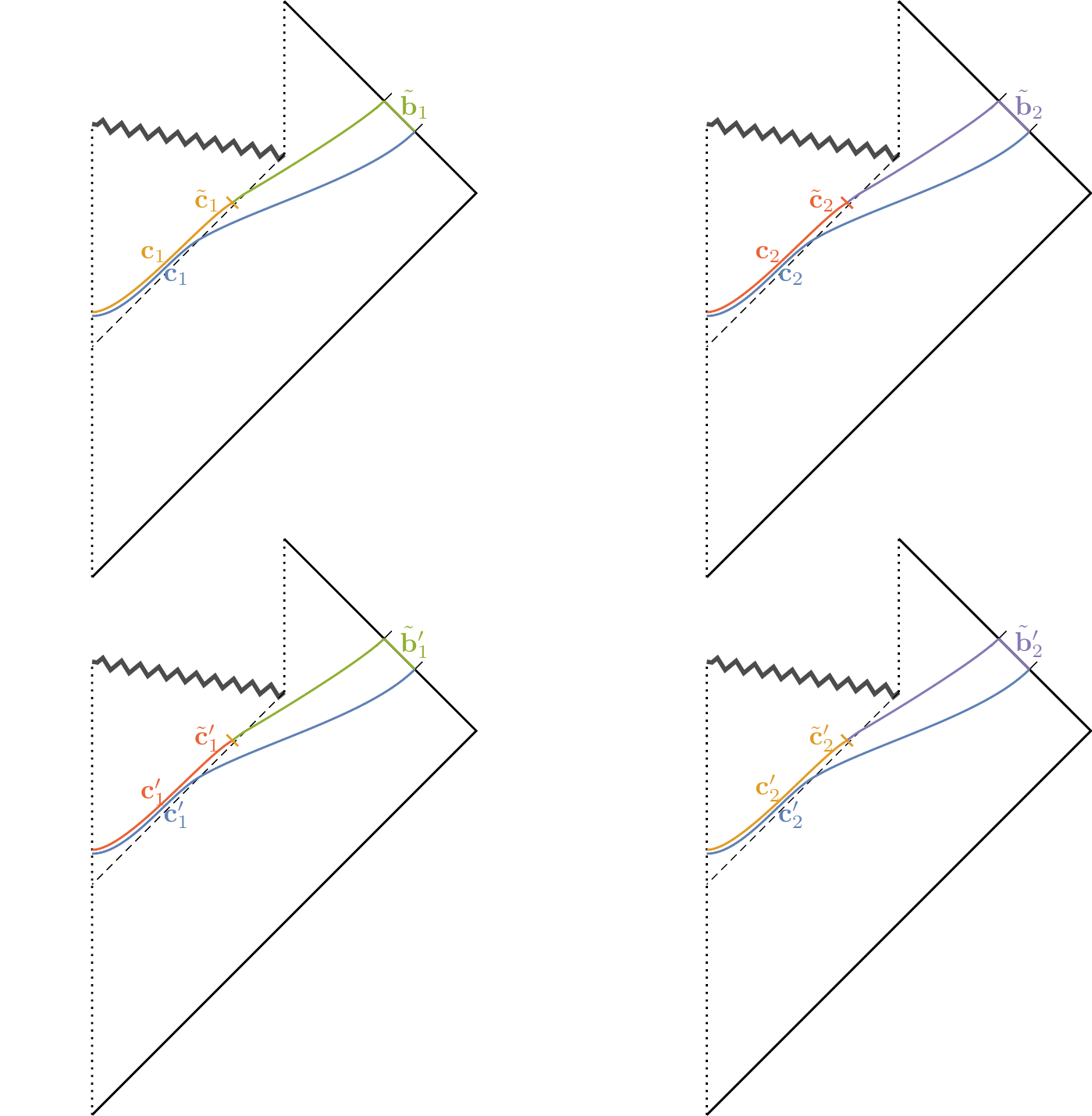}
	\caption{An illustration of a replica wormhole spacetime contributing to the inner product $\llangle \mathbf{c}_1',\mathbf{c}_2'|\mathbf{c}_1,\mathbf{c}_2\rrangle$ in $\hilb_k^{(2)}$.  The left and right diagrams indicate the first and second replica respectively, while the top and bottom represent `ket' and `bra'. The relevant piece of the spacetime comprises the four pieces bounded by the lower Cauchy surfaces $\Sigma_k$ (in blue) and the upper Cauchy surfaces $\Sigma_{k_\island} \cup R$ in yellow, green, purple, red. The path integral on these pieces computes matrix elements of a time-evolution operator $U$ (top) or $U^\dag$ (bottom), with boundary conditions on the lower blue $\Sigma_k$ boundary fixed by the states of interest. Time evolution leaves $\mathbf{c}$ modes unaltered, while creating an entangled state of new interior $\tilde{\mathbf{c}}$ modes and exterior Hawking radiation $\tilde{\mathbf{b}}$ modes (with $\tilde{\mathbf{b}}$ and $\tilde{\mathbf{c}}$ modes summed over). The boundary conditions on the future slices identify fields according to colour. The radiation modes are identified in the obvious manner, setting $\tilde{\mathbf{b}}_1=\tilde{\mathbf{b}}_1'$ (green) and $\tilde{\mathbf{b}}_2=\tilde{\mathbf{b}}_2'$ (purple). The interior modes on the `island' are instead swapped, identifying $\mathbf{c}_1=\mathbf{c}_2'$, $\tilde{\mathbf{c}}_1=\tilde{\mathbf{c}}_2'$ (yellow), and $\mathbf{c}_2=\mathbf{c}_1'$, $\tilde{\mathbf{c}}_2=\tilde{\mathbf{c}}_1'$ (red). The matter path integral on this geometry thus computes the matrix elements $\langle \mathbf{c}_1',\mathbf{c}_2'|\tau_2(\island)|\mathbf{c}_1,\mathbf{c}_2\rangle$ of the swap operator $\tau_2(\island)$ acting on fields on the island, while the gravitational action contributes a factor of $\lambda_\island = e^{-\frac{1}{4G}\operatorname{Area}(\partial\island)}$.  \label{fig:RW}}
\end{figure}
 
 We now briefly explain the replica geometry $\dmanifold$ associated with $\hat{\pi}(\island)$: see figure \ref{fig:RW} and \cite{Marolf:2020rpm,Marolf:2021ghr} for details. Since the operator involves conjugating with time-evolution, $\dmanifold$ is a `timefold' geometry, or an in-in or Schwinger-Keldysh contour (a standard formalism for computing real-time quantities). The matrix elements of the `forward' time-evolution operator $U$ is computed by a path integral over the Lorentzian spacetime bounded by $\Sigma_k$ in the past and $\Sigma_{k_\island}$ in the future, weighted as usual  by $e^{iS}$  with the action $S$. Since we need to conjugate by $U$ we also must compute the inverse $U^\dag$, which we do with a path integral over the same spacetime but  weighted by the complex conjugate action $e^{-iS}$.\footnote{More precisely these path integrals are related by an action of CPT conjugation, which in general may act nontrivially on the fields.} For each of these `forward' and `backward' branches of the spacetime, we have $n$ identical copies, one for each replica.
 
 It remains only to explain the boundary conditions for these $2n$ spacetimes. In the past (on $\Sigma_k$), the boundary conditions are defined by the states $|\psi\rangle$ and $\langle \phi|$ whose inner product we are interested in (so we are computing $\langle\phi|\eta_\dmanifold|\psi\rangle$ as in \eqref{eq:IPconstr1}). In the future (on $\Sigma_{k_\island}$), we act with the permutation operator $\hat{\pi}$ on the island region, before setting field configurations on forward and backward branches equal and integrating over them. This identification and integration of fields effectively stitches the manifold together, giving us a geometric representation of the operator $\hat{\pi}(\island)$. Outside the island, we identify fields on the future slice $\Sigma_{k_\island}$ in the trivial way: the $r$th replica for $U$ is glued to the $r$th replica for $U^\dag$, performing a partial trace over $\bar{\island}$. Within the island region $\island$, we instead identify the $r$th replica on the $U$ spacetime with the $\pi(r)$th replica on the $U^\dag$ spacetime. This gives us a geometric connection between replicas (as long as they are part of the same cycle of $\pi$).

Now that we understand the operators $\eta_\dmanifold$ from the path integral of fluctuating fields on the background $\dmanifold$, we just need to understand the weighting coefficients $\lambda_\dmanifold$ of replica wormholes appearing in \eqref{eq:IPconstr1}. These arise from the gravitational action of $\dmanifold$ itself. It may be surprising that this is nonzero, as the two branches of the timefolds contribute to the path integral measure with equal and opposite actions, as $e^{iS}$ and $e^{-iS}$ respectively. But in fact we should include a \emph{real} contribution to the path integral (corresponding to imaginary Lorentzian action) from the codimension-two surface $\partial\island$. This is a singular locus of the spacetime due to changing the identification along $\tilde{\Sigma}_0$ from trivial to the permutation $\pi$, giving a Lorentzian analogue of a conical excess. In Einstein gravity, this contributes an imaginary action proportional to the area of the surface $\partial\island$. This conclusion follows by defining the action using a Lorentzian analogue of the Gauss-Bonnet theorem applied in a neighborhood of $\partial\island$, or by perturbing the metric to be nonsingular but complex-valued \cite{Louko:1995jw,Witten:2021nzp,Colin-Ellerin:2020mva,Colin-Ellerin:2021jev}.
\begin{equation}\label{eq:RWgravity}
	\text{Gravitational action} \longrightarrow \lambda_\dmanifold = \lambda_\mathcal{I}^{|\pi|},\quad \text{where }  \lambda_\mathcal{I}=e^{-\frac{1}{4G}\operatorname{Area}(\partial\island)}.
\end{equation}
The exponent $|\pi|$ is an integer defining the `size' of the permutation $\pi$. This is given by $|\pi|=n-(\#\text{cycles})$, where $\#\text{cycles}$ is the number of disjoint cycles in $\pi$: in particular, for the cyclic permutation $\pi=\tau_n$ we have a single cycle of length $n$ so $|\tau_n|=n-1$.\footnote{For an alternative definition, construct a graph from $\Sym(n)$ where each vertex corresponds to an element $\pi\in \Sym(n)$, and $\pi_1,\pi_2$ are joined by an edge whenever $\pi_1^{-1}\cdot \pi_2$ is a transposition. Then $|\pi|$ is the distance between $\pi$ and the identity in this graph. In other words, $|\pi|$ gives the minimal number of terms required to write $\pi$ as a product of transpositions.} We will see the same quantity $|\pi|$ also shows up in quantum calculations,  from counting the number of independent sums over states.

Putting this gravitational action together with the permutation operator, the inner product \eqref{eq:IPconstr1} obtained from summing over replica wormholes becomes
\begin{equation}
\begin{gathered}
	\llangle \phi|\psi\rrangle = \langle \phi|\constr|\psi\rangle, \qquad 
	\constr = \id + \sum_{\island
	} \sum_{\substack{\pi\in\Sym(n)\\ \pi\neq \id}} \lambda_\island^{|\pi|} \, \hat{\pi}(\island), \label{eq:IPconstr2}
	\end{gathered}
\end{equation}
where we sum both over nontrivial  permutations $\pi$ and over the location of the island $\island$. This is the structure we will emulate for our bit models.

We conclude this section with some comments on the formula \eqref{eq:IPconstr2} in realistic models of quantum gravity.

The operator $\hat{\pi}(\island)$ does not make sense in continuum quantum field theory, since its matrix elements will be UV divergent (as is familiar from entropy in QFT). However, this divergence is compensated for by the renormalisation of the gravitational action \cite{Susskind:1994sm,Jacobson:1994iw,Frolov:1996aj}: the area in \eqref{eq:RWgravity} is corrected by counterterms, and we expect these to precisely cancel the divergence coming from $\hat{\pi}(\island)$ so that this particular combination is finite and independent of renormalisation scheme. The combination of area \eqref{eq:RWgravity} and entropy \eqref{eq:pientropy} is the `generalised entropy'. With this in mind, we can interpret $G$ as the IR Newton's constant and the QFT matrix element as regulated on some IR scale.

The formula \eqref{eq:IPconstr2} represents a sum over some family of manifolds, but at no point did we say that we are summing over saddle-points. Indeed, the existence and location of a saddle-point may depends on precisely what quantity we would like to compute. Furthermore, the purely Lorentzian replica wormhole geometries can never be saddle-points, because the equations of motion are not satisfied on the singular splitting surface $\island$. Our attitude is that the sum (or integral) over islands $\island$ in \eqref{eq:IPconstr2} represents an off-shell integration over some contour in configuration space, and for any particular calculation it may be convenient to deform this contour to pass through a saddle-point (often expected to involve a complex metric). See \cite{Marolf:2022ybi} and further comments and questions in section \ref{ssec:LorentzianPI}.

\subsection{Replica wormhole bit models}\label{sec:RWmodels}

We are now ready to incorporate replica wormholes into the bit model of section \ref{sec:singleBH}. These model the Hilbert spaces of multiple black holes, describing the state on $n$ copies of a surface $\Sigma_k$ in a Hilbert space $\hilb_k^{(n)}$.

Following the idea outlined in section \ref{sec:NPIP}, our starting point is a `perturbative Hilbert space' $\hilb_\mathrm{pert}$, which we construct in the conventional manner from the $n$-fold tensor product of spaces $\hilb_k$ as described in section \ref{sec:singleBH}:
\begin{equation}
\begin{gathered}
	\hilb_\mathrm{pert} =\hilb_k^{\otimes n} \text{ spanned by } |\mathbf{c}_1;\ldots;\mathbf{c}_n\rangle = |\mathbf{c}_1\rangle\otimes\cdots\otimes|\mathbf{c}_n\rangle, \\
	\langle \mathbf{c}_1';\ldots;\mathbf{c}_n'|\mathbf{c}_1;\ldots;\mathbf{c}_n\rangle = \delta_{\mathbf{c}_1'\mathbf{c}_1}\cdots \delta_{\mathbf{c}_n'\mathbf{c}_n} \qquad \text{(no replica wormholes)},
\end{gathered}
\end{equation}
where each $|\mathbf{c}_r\rangle \in \mathcal{H}_k$ is described by a $k$-tuple of bits $\mathbf{c}_r = (c_{1,r},c_{2,r},\ldots,c_{k,r})$, with the replica index $r$ running from $1$ to $n$. The dynamics similarly factorises in such a model: the evolution from $\mathcal{H}_{k-1}^{\otimes n}$ to $\mathcal{H}_{k}^{\otimes n}$ is given by $n$ copies $U_k^{\otimes n}$ of the  time-evolution operator $U_k$ defined in \eqref{eq:U}, acting separately on each factor.

The physical Hilbert space $\hilb_\mathrm{phys}=\hilb_k^{(n)}$ retains the same set of spanning states labelled by $n$ $k$-tuples of bits $\mathbf{c}_r$ and the same factorising dynamics. The only modification we make is to the inner product, which (as in section \ref{sec:NPIP}) we will write with doubled kets $|\ \cdot\ \rrangle$, so we have
\begin{equation}\label{eq:IPeta}
\begin{gathered}
	\hilb_\mathrm{phys} = \mathcal{H}_k^{(n)} \text{ spanned by } |\mathbf{c}_1;\ldots;\mathbf{c}_n\rrangle , \\
	\llangle \mathbf{c}_1';\ldots;\mathbf{c}_n'|\mathbf{c}_1;\ldots;\mathbf{c}_n\rrangle = \langle\mathbf{c}_1';\ldots;\mathbf{c}_n'| \eta_{k,n}|\mathbf{c}_1;\ldots;\mathbf{c}_n\rangle. 
\end{gathered}
\end{equation}
To avoid confusion, we have made explicit the dependence of the matrix of inner products $\eta_{k,n}$ on $n$ and $k$.

Following our discussion of replica wormholes in section \ref{sec:RWIP}, we define $\eta_{k,n}$ as a weighted sum \eqref{eq:IPconstr2} over permutation operators $\hat{\pi}(\island)$, with $\pi\in\Sym(n)$  acting on `islands' $\island$. Here, the `island' $\island$ is now simply a set of bits $\{1,2,\ldots,k_\island\}$ for some $k_\island$ on which the permutation acts.\footnote{A straightforward  generalisation allows $\island$ to be any subset of indices, but the resulting extra islands aren't important in any case we know of so we exclude them for simplicity.} This models a partial Cauchy surface covering the interior of the black hole at time $t_{k_\island}$, bounded by the event horizon at that ingoing time. If $k_\island\leq k$, then $\hat{\pi}(\island)$ simply acts by the permutation $\pi$ on the first $k_\island$ indices of the $n$ replicas. To write this explicitly as in \eqref{eq:permop}, we split $\mathbf{c}_r =  (\hat{\mathbf{c}}_r,\tilde{\mathbf{c}}_r) $ for each $r=1,\ldots,n$, where $\hat{\mathbf{c}}_r$ contains the first $k_\island$ bits and $\tilde{\mathbf{c}}_r$ the remaining $k-k_\island$, and then
\begin{equation}
	\hat{\pi}(\island) | \hat{\mathbf{c}}_1,\tilde{\mathbf{c}}_1;\ \cdots\ ;\hat{\mathbf{c}}_n,\tilde{\mathbf{c}}_n\rangle = | \hat{\mathbf{c}}_{\pi^{-1}(1)},\tilde{\mathbf{c}}_1 ;\ \cdots\ ;\hat{\mathbf{c}}_{\pi^{-1}(n)},\tilde{\mathbf{c}}_n\rangle.
\end{equation}
However, we may also take $k_\island>k$, so it acts on a larger number of bits. This means that we must apply the evolution operators $U$ at least $k-k_\island$ times on each copy to evolve to a Cauchy slice containing the island, then act with the permutation, and finally trace over the additional bits (both interior partners and radiation) produced in the evolution. We write this out in detail and compute its effect in equation \eqref{eq:piisland} in a moment. This feature is crucial for unitarity of the evolution.

With these ingredients, we finally can give an explicit definition of the physical inner product:
\begin{gather}
	\eta_{k,n} = \id + \sum_{k_\island =1
	}^K \sum_{\substack{\pi\in\Sym(n)\\ \pi\neq \id}} \lambda_{\island}^{|\pi|} \, \hat{\pi}(\island).\label{eq:constraintisland}
\end{gather}
This defines a class of models since we still have the freedom to define the weightings $\lambda_\island$, which we recall have the gravitational interpretation of exponential area suppressions $\exp\left(-\frac{A(\partial\island)}{4G}\right)$. We will define weights for a couple of concrete models in sections \ref{sec:PS} and  \ref{sec:RWmodel}.

Our treatment of the islands with $k_\island>k$ was rather implicit above, so we now spell out the details and calculate the consequences of the definition, computing the matrix elements of $\hat{\pi}(\island)$ with  $k_\island>k$:
\begin{align}
\langle \mathbf{c}_1';\ldots;\mathbf{c}_n'| &\hat{\pi}(\island) |\mathbf{c}_1;\ldots;\mathbf{c}_n\rangle \nonumber\\
	 &:=  \left(\langle \mathbf{c}_1'|({U^\dag})^{k_\island-k}\otimes\cdots\otimes\langle\mathbf{c}_n' |({U^\dag})^{k_\island-k}\right) \hat{\pi} \left(U^{k_\island-k}|\mathbf{c}_1\rangle\otimes \cdots\otimes U^{k_\island-k}|\mathbf{c}_n\rangle\right)\nonumber \\
	&= \sum_{\tilde{\mathbf{b}}_r,\tilde{\mathbf{b}}_r',\tilde{\mathbf{c}}_r,\tilde{\mathbf{c}}_r'}  \langle \mathbf{c}_1',\tilde{\mathbf{c}}_1';\ \cdots\ \mathbf{c}_n',\tilde{\mathbf{c}}_n' | \hat{\pi} |\mathbf{c}_1,\tilde{\mathbf{c}}_1;\ \cdots\ ;\mathbf{c}_n,\tilde{\mathbf{c}}_n\rangle  \nonumber \\
	&\mkern 150mu \times \bar{\psi}_{\tilde{\mathbf{b}}_1'\tilde{\mathbf{c}}_1'} \psi_{\tilde{\mathbf{b}}_1\tilde{\mathbf{c}}_1} \langle \tilde{\mathbf{b}}_1'|\tilde{\mathbf{b}}_1\rangle_R \ \cdots\  \bar{\psi}_{\tilde{\mathbf{b}}_n'\tilde{\mathbf{c}}_n'} \psi_{\tilde{\mathbf{b}}_n\tilde{\mathbf{c}}_n}\langle \tilde{\mathbf{b}}_n'|\tilde{\mathbf{b}}_n\rangle_R \nonumber\\
	&=\frac{1}{2^{n(k_\island-k)}}\sum_{\tilde{\mathbf{c}}_r}  \delta_{\mathbf{c}_1'\mathbf{c}_{\pi^{-1}(1)}} \delta_{\tilde{\mathbf{c}}_1\tilde{\mathbf{c}}_{\pi^{-1}(1)}}\  \cdots \ \delta_{\mathbf{c}_n'\mathbf{c}_{\pi^{-1}(n)}} \delta_{\tilde{\mathbf{c}}_n\tilde{\mathbf{c}}_{\pi^{-1}(n)}} \nonumber\\
	&= 2^{-|\pi|(k_\island-k)} \delta_{\mathbf{c}_1'\mathbf{c}_{\pi^{-1}(1)}}  \cdots \ \delta_{\mathbf{c}_n'\mathbf{c}_{\pi^{-1}(n)}} \label{eq:piisland}  \,.
\end{align}
Here $\tilde{\mathbf{c}}_r$ and $\tilde{\mathbf{c}}'_r$ are $(k_\island-k)$-tuples of bits describing the Hawking partners created between the slices $\Sigma_k$ and $\Sigma_{k_\island}$, and $\tilde{\mathbf{b}}_r$, $\tilde{\mathbf{b}}_r'$ similarly describe the radiation with which these are entangled. The inner product on the radiation sets $\tilde{\mathbf{b}}_r=\tilde{\mathbf{b}}_r'$. The wavefunctions $\psi_{\tilde{\mathbf{b}}_r\tilde{\mathbf{c}}_r}$ are products of $k_\island-k$ terms $\psi_{\tilde{b}_{r,k+1}\tilde{c}_{r,k+1}} \!\cdots \,\ \psi_{\tilde{b}_{r,k_\island}\tilde{c}_{r,k_\island}}$, one for each bit of radiation from each replica.

The upshot is
\begin{equation}
	\hat{\pi}(\island) = 2^{-|\pi|(k_\island-k)} \hat{\pi}\quad \text{for $k_\island\geq k$ acting on $\hilb_k^{\otimes n}$.}
\end{equation}
The interesting result of this calculation is the suppression $2^{-|\pi|(k_\island-k)}$ from tracing out the $k_\island-k$ bits of radiation. At a technical level, this results from the difference between the $n$ normalising factors of $2^{k_\island-k}$ (counting all possible states of the radiation) and the smaller number of independent sums over $\tilde{\mathbf{c}}_r$ resulting from identifications $\delta_{\tilde{\mathbf{c}}_r\tilde{\mathbf{c}}_{\pi(r)}}$ between replicas in the same cycle of the permutation. Not coincidentally, the result is controlled by the same power of $|\pi|$ characteristic of the permutation that appeared from geometrical considerations in \eqref{eq:RWgravity}. These two pieces combine into powers of $\lambda_\island 2^{-(k_\island-k)}=e^{-S_\mathrm{gen}}$, where $S_\mathrm{gen}$ is interpreted as a `generalised entropy': a sum of the area of $\partial\island$ and matter entropy on $\island$.

For the inner product between pure product states (such as the basis states $| \mathbf{c}_1;\ldots;\mathbf{c}_n\rangle$ in \eqref{eq:piisland}), only the later radiation (bits $c_i$ with $i>k$) contributes to this `generalised entropy'.  For calculations involving mixed states on $\Sigma_k$ (such as the interior state created by Hawking evaporation up to this time), the additional matter entropy will tend to contribute to the $S_\mathrm{gen}$ suppression similarly due to the same permutation effect.

\subsection{Some properties of the replica wormhole inner product}\label{sec:properties}

We now note some key properties of the replica wormhole inner product defined above.

First, consider a semiclassical regime where $n$ is not too large, and for every island $\island$ either $\lambda_\island\ll 1$ or $k_\island-k \gg 1$: that is, all possible islands appear when the black hole has large area in Planck units, or if not (which allows for islands a long time in the future when the black hole is close to complete evaporation) they appear after many more bits of radiation are emitted. Then, the coefficients of the non-identity terms in $\constr$ are small, so (assuming additionally that there are not too many terms to add up to something large) the modifications to the usual tensor product inner product are small: for most questions, the corrections are negligible. This is important so that we retain normal local physics such as a smooth horizon, discussed in more detail in section \ref{sec:smooth}. Note that this can fail if we take $n$ to be exponentially large, which is an interesting regime that we comment on in the discussion of section \ref{sec:disc}.

Next, $\constr_{k,n}$ is manifestly Hermitian  on  $\hilb_k^{\otimes n}$, and if the coefficients $\lambda_\island$ are small it is positive semi-definite. These properties are essential so that \eqref{eq:constraintisland} really defines an inner product. Positive-definiteness is not completely obvious and we do not know of a simple condition to guarantee it (for large $\lambda_\island$ it can certainly fail), but the above considerations (and more careful calculations in section \ref{sec:smooth}) guarantee that it cannot fail in the semiclassical regime that concerns us  in this paper.

The marginal case where $\constr$ is truly semi-definite with a nontrivial kernel is particularly interesting. In that case, $\hilb_k^{(n)}$ is not the same as  $\hilb_k^{\otimes n}$ even as a vector space, since we must quotient by the `null states': $\hilb_k^{(n)}\simeq (\hilb_k)^{\otimes n}/\ker\constr$. That is, some apparently distinct states are in fact `gauge equivalent' in the sense of \cite{Marolf:2020xie,Jafferis:2017tiu}. However, again from the discussion above this cannot occur in the semiclassical regime. We will not emphasise this aspect in the present paper, but will explore the consequences and interpretation in future work (with a few comments in section \ref{ssec:null}).

Finally, the unitarity of time evolution is exactly preserved by the model. That is, $U$ as defined in \eqref{eq:U} continues to preserve the overlap of states with the physical inner product: $U^{\otimes n}$ is an isometry from $\hilb_{k-1}^{(n)}$ to $\hilb_{k}^{(n)}\otimes \hilb_{R_{\{k\}}}^{\otimes n}$ with respect to the inner products $\llangle\cdot|\cdot\rrangle$ on $\hilb_{k}^{(n)}$ and the usual inner product on the radiation $\hilb_{R_{\{k\}}}$. In particular, if there are null states it well-defined on the quotient space $(\hilb_k)^{\otimes n}/\ker\constr$, since $U$ preserves $\ker\constr$. This is in fact manifest from the way that we defined $\hat{\pi}(\island)$, since we can always choose to compute the inner product by first evolving using $U$ to the final slice $\Sigma_K$ (with $K$ the largest possible value of $k_\island$). It also follows straightforwardly from direct calculation. Note that for this to hold it is crucial that islands contribute to the inner products at all times, even though they are exponentially suppressed when $k_\island-k\gg 1$. This aspect was neglected in \cite{Guo:2021blh}.

\subsection{The Polchinski-Strominger model}\label{sec:PS}

While the full replica wormhole bit model discussed above is quite concrete and conceptually simple, it still a little complicated since it involves the sum over all possible islands, and it requires us to make a choice of the weights $\lambda_\island$. But the most important features of the model are present for a simplified version. For this, we allow only one possible island $\island=\{1,2,\ldots,K\}$ consisting of the entire black hole interior after complete evaporation, with trivial weighting $\lambda_\island=1$ (all other $\lambda_\island$ are zero).  We call this the Polchinski-Strominger (PS) model due to the similarity with the proposal of \cite{Polchinski:1994zs}, as explained in \cite{Marolf:2020rpm}.



In this simplified model, the inner product \eqref{eq:constraintisland} on  $\hilb^{(n)}_k$ is given by
\begin{equation}
	\llangle  \mathbf{c}_1';\ldots;\mathbf{c}_n' |\mathbf{c}_1;\ldots;\mathbf{c}_n\rrangle =\sum_{\pi\in \Sym(n)} 2^{-|\pi|(K-k)} \delta_{\mathbf{c}_1'\mathbf{c}_{\pi^{-1}(1)}}  \cdots \ \delta_{\mathbf{c}_n'\mathbf{c}_{\pi^{-1}(n)}} \label{eq:PSIP}  \,.
\end{equation}
Alternatively, we can express this in terms of the operator $\eta_{k,n}$ (now making the dependence on $k$ and $n$ explicit) giving the physical inner product:
\begin{equation}\label{eq:PSeta}
	\eta_{k,n} = \sum_{\pi\in\Sym(n)} 2^{-|\pi|(K-k)} \hat{\pi},
\end{equation}
where $\hat{\pi}$ here is the permutation operator acting on $\hilb_{k}^{\otimes n}$. We can also write this as
\begin{equation}
	\eta_{k,n} = \frac{1}{2^{n(K-k)}} \sum_{\pi\in\Sym(n)}\Tr_{K-k} (\hat{\pi}_K),
\end{equation}
where $\hat{\pi}_K$ is the permutation operator acting on $K$ bits (the entire interior), and $\Tr_{K-k}$ is a partial trace over the last $K-k$ bits (interpreted as tracing out the future Hawking radiation).

On the final slice $k=K$, this becomes simply the Bose inner product summing over all permutations, $\hilb^{(n)}_K = \Sym^n(\hilb_K)$. Equivalently, we allow only symmetric wavefunctions of the $n$ copies: $\eta_{K,n} = \sum_\pi \hat{\pi}_K = n! P_{\Sym}$ is $n!$ times a projector $P_{\Sym}$ onto these symmetric wavefunctions. This means we treat the black hole interiors like bosons, as indistinguishable closed universes. If we suppose that the black hole interior splits off from the exterior at the end of the evaporation so that we can treat $\Sigma_K$ as a Cauchy surface of a closed universe as in figure \ref{fig:finalSlice}, then this symmetrisation is a simple (albeit non-perturbative) consequence of gauging diffeomorphisms: specifically, the diffeomorphisms of space which permute the closed universes.

In fact, versions of this model have appeared recently in two different contexts. The first is JT gravity with end-of-the-world branes \cite{Penington:2019kki} (and similarly the simple topological model of \cite{Marolf:2020xie}). In that model the inner product on internal brane states takes essentially the same form (once the spectrum of the Hamiltonian is fixed), due to a very similar sum over topologies. The second is the non-dynamical model of \cite{Akers:2022qdl} if we take the full ensemble of models seriously (rather than picking just one representative); computing moments of of the random unitary in that model reduces to a sum of permutations which mirrors our sum over wormhole topologies. Comparison with these models will be easier later using the language of superselection sectors, so we defer a full comparison to the end of section \ref{sec:factorising}.

\subsection{A more realistic model}\label{sec:RWmodel}

While the Polchinski-Strominger model captures most qualitative features of interest (as we will explore in the sequel), it is too simple for some purposes. Some of its shortcomings are discussed in section 4.4 of \cite{Marolf:2020rpm}. Here we define a concrete model which includes a family of replica wormholes, and hence reproduces more interesting features of gravity.

To motivate this, we note one unrealistic feature of the Polchinski-Strominger model, which we will see explicitly in the next section. Namely, it gives the entropies expected of an adiabatic evaporation, in the sense that the total coarse-grained entropy (of black hole and radiation) is constant, saturating the second law of thermodynamics. In freely evaporating black holes, this does not hold: there is thermal entropy production. The thermal entropy in question is measured by the `generalised entropy', namely the area of the horizon plus the thermodynamic (coarse-grained) entropy outside the black hole:
\begin{equation}
	S_\mathrm{gen} = \frac{A}{4G_N} +S_\mathrm{out}.
\end{equation}
The statement that this increases is the generalised second law (GSL), a quantum version of Hawking's area theorem \cite{Bekenstein:1974ax}. If there is no extra scale in the problem (such as for black holes emitting massless Hawking radiation in an asymptotically flat spacetime), $S_\mathrm{gen}$ strictly increases in a simple manner: for each unit that $\frac{A}{4G_N}$ decreases as the black hole shrinks, the radiation emitted in the process has an entropy $r$ times larger for some fixed $r>1$. This ratio was computed in examples by Page \cite{Page:1976df} (e.g.~for gravitons and photons emitted from a Schwarzschild black hole, we have $r\approx 1.48$).

We define a one-parameter family of models designed to reproduce this feature. In our model, the coarse-grained entropy of Hawking radiation increases by a constant $\log 2$ at each time-step (since the state is maximally mixed on a single qubit). So the area should decrease similarly at a constant rate, going to zero at the end of evaporation. Since $\lambda_\island$ is interpreted as exponential in the area, we should choose
\begin{equation}
	\lambda_{\island} = x^{k_\island-K} \qquad (1<x<2).
\end{equation}
This means that the black hole entropy $\frac{A}{4G_N}$ decreases by a constant $\log x$ at each time step. We require $x>1$ so that the black hole actually shrinks, and $x<2$ to satisfy the GSL with the interpretation $\frac{A}{4G}=(K-k)\log x$. In terms of the ratio $r$ above, we have $x=2^{\frac{1}{r}}$. For $x<2$, the islands in the late stages of evaporation (small $K-k_\island$) are exponentially suppressed in the semiclassical regime of large $K-k$. In contrast, for $x>2$ the sum over islands will tend to be dominated by the final slice $k_\island=K$, so we effectively recover the Polchinski-Strominger model above (which we recover exactly with $x\to\infty$).

\section{The Page curve}\label{sec:Page}

With our model in hand, we can now demonstrate its first key feature: the Page curve for the entropy of Hawking radiation.

\subsection{Operationally defined entropy}

We take the perspective of \cite{Marolf:2021ghr} (more details in \cite{Marolf:2020rpm}) that the physically relevant entropy is defined operationally: it is deduced from the results of observations on the radiation. In particular, since von Neumann entropy is not linear in the density matrix, without additional information it can never be determined using a single copy of the system. Rather, it requires joint measurements on several copies (or replicas).

With this in mind, we consider an experiment where we create $n$ identical `replica' black holes  and allow them to partially evaporate, each emitting $k$ bits of Hawking radiation. This gives us a pure state on $\hilb_k^{(n)}\otimes \hilb_{R_k}^{\otimes n}$: the factor  $\hilb_k^{(n)}$ describes the state of the remaining $n$ black holes, and  each $\hilb_{R_k}$ describes the radiation from one of the replicas:\footnote{This state is not quite normalised, though its norm is exponentially close to unity in the semiclassical limit so we can disregard this for our purposes. A slight refinement for normalisation is explained in section \ref{ssec:norms}.}
\begin{equation}
	\sum_{\mathbf{b}_r,\mathbf{c}_r}\psi_{\mathbf{b}_1\mathbf{c}_1}\cdots \psi_{\mathbf{b}_n\mathbf{c}_n} |\mathbf{c}_1;\ldots;\mathbf{c}_n\rrangle_{\hilb_k^{(n)}} \otimes |\mathbf{b}_1\rangle_{R_k}\otimes\cdots \otimes |\mathbf{b}_n\rangle_{R_k} \,.
\end{equation}
 Tracing out the black holes (using the physical inner product on $\hilb_k^{(n)}$) leaves us with a state on $\hilb_{R_k}^{\otimes n}$ with density matrix $\rho_{R_k}^{(n)}$:
 \begin{equation}\label{eq:rhoRkn}
	\langle\mathbf{b}_1;\ldots;\mathbf{b}_n|\rho_{R_k}^{(n)}|\mathbf{b}'_1;\ldots;\mathbf{b}'_n\rangle=\sum_{\mathbf{c}_r,\mathbf{c}_r'}\psi_{\mathbf{b}_1\mathbf{c}_1}\bar{\psi}_{\mathbf{b}_1'\mathbf{c}_1'}\cdots \psi_{\mathbf{b}_n\mathbf{c}_n}\bar{\psi}_{\mathbf{b}_n'\mathbf{c}_n'} \llangle\mathbf{c}_1';\ldots;\mathbf{c}_n'|\mathbf{c}_1;\ldots;\mathbf{c}_n\rrangle.
\end{equation}
 The results of all possible measurements on the radiation are encoded by the expectation values of operators on $\hilb_{R_k}^{\otimes n}$ in that state.

The specific operators we are interested in are cyclic permutation operators $\hat{\tau}_n$ acting on the radiation, with $\tau_n$ the cyclic permutation defined by $\tau_n(r)=r+1$ for $r=1,2,\ldots,n-1$ and $\tau_n(n)=1$. This operator is defined similarly to the permutation operators we saw in section \ref{sec:RWIP} acting on an island, but differ in interpretation since operators on the radiation are `external' (things we choose to measure in a region where quantum gravitational effects are negligible), while the island permutation operators are `dynamical' (things we must sum over to compute an amplitude). These operators have a close relationship to fine-grained entropy, which we saw already in \eqref{eq:pientropy}. This motivates us to define a `swap R\'enyi entropy' via their expectation values:
\begin{equation}
	S^\mathrm{swap}_n(R_k) = -\frac{1}{n-1}\log \Tr(\rho_{R_k}^{(n)} \hat{\tau}_n).
\end{equation}
The name is explained by two things. First, the $n=2$ version $\hat{\tau}_2$ is the `swap operator', and measuring this is the `swap test' to distinguish mixed from pure states \cite{buhrman2001quantum,Hayden:2007cs}; we are generalising this to any replica number $n$. Second, in the case $\rho^{(n)} = \rho^{\otimes n}$ the swap entropy gives precisely the $n$th R\'enyi entropy of $\rho$. Finally, since the von Neumann entropy is the $n\to 1$ limit of the R\'enyi entropy, we define a `swap von Neumann entropy'
\begin{equation}
	S^\mathrm{swap}(R_k) = \lim_{n\to 1} S^\mathrm{swap}_n(R_k)
	\end{equation}
	(subject to the usual replica trick comments about continuing from integer $n$).

Now, using the state of radiation \eqref{eq:rhoRkn} we can write the expectation value of a permutation operator directly in terms of the physical inner product on $\hilb_k^{(n)}$. The form of the result is particularly simple in our model where the state $\psi_{\mathbf{b}\mathbf{c}}$ of Hawking radiation is maximally mixed:
 \begin{equation}
 \begin{aligned}
 	\Tr(\rho_{R_k}^{(n)}\hat{\pi}) &= \frac{1}{2^{nk}}\sum_{\mathbf{c}_r} \llangle\mathbf{c}_{\pi^{-1}(1)};\ldots;\mathbf{c}_{\pi^{-1}(n)}|\mathbf{c}_1;\ldots;\mathbf{c}_n\rrangle \\
 	&= \frac{1}{2^{nk}}\Tr(\hat{\pi}^\dag \eta_{k,n}) \qquad (\text{trace taken in }\hilb_k^{\otimes n}).
 \end{aligned}
\end{equation}
In the second line we have used the form \eqref{eq:IPeta} of the physical inner product, as the matrix elements of an operator $\eta_{k,n}$ in the tensor product Hilbert space $\hilb_k^{\otimes n}$. For a model with a general state $\psi_{\mathbf{b}\mathbf{c}}$, this result would be modified by replacing the normalising factor $2^{-nk}$ with $n$ insertions of the Hawking partner density matrix (with matrix elements $\sum_\mathbf{b} \psi_{\mathbf{b}\mathbf{c}}\bar{\psi}_{\mathbf{b}\mathbf{c}'}$), one for each replica.

\subsection{Entropies in the Polchinski-Strominger model}

We begin with the simple case of the Polchinski-Strominger model defined in section \ref{sec:PS}, which has only one sort of island. Using \eqref{eq:rhoRkn} and the expression \eqref{eq:PSeta} for the physical inner product, we find
\begin{equation}
\begin{aligned}
	\Tr(\rho_{R_k}^{(n)}\hat{\tau}_n) &= \frac{1}{2^{nk}}\sum_{\pi\in\Sym(n)} 2^{-|\pi|(K-k)} \Tr(\hat{\tau}_n^\dag\hat{\pi}) \\
	&= \sum_{\pi\in\Sym(n)} 2^{-|\tau_n^{-1}\circ\pi|k-|\pi|(K-k)}.
	\end{aligned}
\end{equation}
The trace in the second line comes from counting the number of independent sums in the trace, similarly to \eqref{eq:piisland}.

To see the main idea, let's look at the simplest case $n=2$. There are only two terms in the sum, coming from $\pi=\id$ and $\pi = \tau_2$:
\begin{equation}
	\Tr(\rho_{R_k}^{(2)}\hat{\tau}_2) = 2^{-k} + 2^{-(K-k)}.
\end{equation}
In the semiclassical limit where both $k$ and $K-k$ are large, one term will dominate over the other (except close to the `Page time' $k=\frac{K}{2}$ where they exchange dominance).
\begin{equation}
	S^\mathrm{swap}_2(R_k) \sim \min\{k,K-k\} \log 2.
\end{equation}

The case for general $n$ is similar. If $k\ll \frac{K}{2}$, the sum over permutations is dominated by the term which minimises $|\pi|$, namely $\pi = \id$. If $k\gg \frac{K}{2}$, the largest term comes from minimising $|\tau_n^{-1}\circ\pi|$, so the permutation $\pi = \tau_n$ matches the permutation we are measuring. This gives the same swap entropy for all $n$:
\begin{equation}
	S^\mathrm{swap}_n(R_k) \sim \min\{k,K-k\} \log 2.
\end{equation}
Close to the transition ($k \approx\frac{K}{2}$), a larger set of terms (those with $|\tau_n^{-1}\circ\pi| + |\pi| = n-1$, which are labelled by non-crossing partitions) are relevant, giving interesting corrections which smooth out the transition \cite{Penington:2019kki,Marolf:2020vsi}.

\subsection{A quantum extremal surface formula}\label{sec:QES}

The idea for  the general model is similar, with the extra interesting feature that $\eta$ involves a sum over the possible islands $\island$ as well as the permutation $\pi$. Our formula \eqref{eq:rhoRkn} becomes
\begin{equation}\label{eq:swapRWsum}
\begin{gathered}
	\Tr(\rho_{R_k}^{(n)}\hat{\tau}_n) = \frac{1}{2^{nk}}\Tr(\hat{\tau}_n^\dag ) + \frac{1}{2^{nk}}\sum_{\island
	} \sum_{\substack{\pi\in\Sym(n)\\ \pi\neq \id}} \lambda_\island^{|\pi|} \, \Tr(\hat{\tau}_n^\dag \hat{\pi}(\island) ).
	\end{gathered}
\end{equation}
The trace is computed similarly to above, with two cases depending on whether the island appears at a later time ($k_\island\geq k$) or an earlier time ($k_\island\leq k$). The first case is identical to the PS model discussed above, with $K$ replaced by $k_\island$. The second case is more interesting: the permutation operator $\hat{\tau}_n^\dag \hat{\pi}(\island)$ acts as $\tau_n^{-1}\circ \pi$ on the $k_\island$ bits making up the island, but only as $\tau_n^{-1}$ on the remaining $k-k_\island$ bits where  $\hat{\pi}(\island)$ does nothing. The result is
\begin{equation}
	\frac{1}{2^{nk}}\Tr(\hat{\tau}_n^\dag \hat{\pi}(\island) ) = \begin{cases}
		2^{-|\tau_n^{-1}\circ\pi|k-|\pi|(k_\island -k)} & k_\island\geq k \\
		2^{-|\tau_n^{-1}\circ \pi| k_\island-|\tau_n^{-1}|(k-k_\island)} & k_\island\leq k
	\end{cases}.
\end{equation}
In addition, we have an `area' term $\lambda_\island^{|\pi|}$.

As for the PS model, the most important island contributions come from the terms  $\pi = \tau_n$ in the sum over permutations. These contribute to \eqref{eq:swapRWsum} as
\begin{equation}\label{eq:Sgen}
\begin{gathered}
	\left(\lambda_\island^{-1}  2^{|k-k_\island|}\right)^{-(n-1)} = e^{-(n-1)S_\mathrm{gen}(\island \cup R_k)}, \\
	\text{where }\quad S_\mathrm{gen}(\island \cup R_k) = \log\lambda_\island^{-1} + |k-k_\island|\log 2.
\end{gathered}
\end{equation}
We have written this in terms of a `generalised entropy': the `area' $\log\lambda_\island^{-1}$ of the island, plus entropy of matter (using the single copy state) on $R_k\cup\island$. For the case $k_\island \leq k$, the first $k_\island$ bits of radiation are purified by the island so don't contribute, leaving the entropy of $k-k_\island$ later bits of radiation; for  $k_\island \leq k$ we similarly count the next $k_\island-k$ Hawking interior modes. The first term in \eqref{eq:swapRWsum} can be regarded as a special case of this with $k_\island=0$ and no area contribution (so $\lambda_\island=1$).

In the case that the expectation value is dominated by the largest term in the sum, we find a `quantum extremal surface formula' \cite{Engelhardt:2014gca} or `island formula':
\begin{equation}
	S^\mathrm{swap}_n(R_k) \sim \min_\island S_\mathrm{gen}(\island \cup R_k).
\end{equation}
A `quantum extremal surface' (QES) is the boundary $\partial\island$ of an island  which locally minimises $S_\mathrm{gen}$, while the global minimum computes the entropy.

For models where the area $\log\lambda_\island^{-1}$ is decreasing, the  only local minimum will occur for $k=k_\island$ (assuming the `GSL' that the  area can not decrease faster than the entropy $\log 2$ emitted per time step; see section \ref{sec:RWmodel}). In that case, we have
\begin{equation}
	S^\mathrm{swap}_n(R_k) \sim \min\{ k \log 2, \log\lambda_\island^{-1} \},
\end{equation}
where the first term comes from the `empty island' term $\pi=\id$ in \eqref{eq:swapRWsum}, and the second from the island $k_\island=k$. In fact, if the ratio of  $\lambda_\island$ at successive values of $k_\island$ is of order unity, terms with $k-k_\island$ of order one will contribute at the same order, but summing over these gives rise only to order one corrections to $ S^\mathrm{swap}$. We can interpret this as an uncertainty of the location of the QES in the direction along the horizon by a time of order $\beta$.

Our model is insufficiently sophisticated to capture correctly more detailed properties in realistic examples. We have $k_\island=k$ instead of the location of  $\partial\island$ trailing by a scrambling time (though perhaps this is a matter of interpretation). Our result for the entropies is also independent of $n$ (a `flat spectrum'), an indication that our model is analogous to a `fixed area state' \cite{Dong:2018seb} where we can ignore backreaction from the splitting surface $\island$. Our island formula involves only a minimum (rather than a local extremum followed by global minimum), because the model does not include fluctuations of the surface in timelike directions. We comment on these issues in section \ref{ssec:LorentzianPI}.

\subsection{The true von Neumann entropies}

We have shown above that operationally defined `swap entropy' follows a Page curve in our model. But if we compute the actual von Neumann entropy of the state $\rho_{R_k}^{(n)}$ of $n$ copies of radiation, we will find that it is  close to the value $n k$ of a maximally mixed state!
%
%
 This is our version of the `state paradox' (as termed in \cite{Bousso:2020kmy}): the entropy computed using islands is different from the manifest entropy of the radiation density matrix.

The resolution in our model is clear. The swap entropy deduced by experiments on the radiation only matches  the `true' entropy under the assumption that identical black holes produce identical, uncorrelated states of radiation. This fails: the replica wormholes correlate the radiation from different replicas.

Since the `swap' entropy deduced by measuring permutations is not the same as the von Neumann entropy, one might worry that other measurements might be incompatible with the Page curve, or imply a different result for the entropy. One might also worry that measurements between widely separated black holes might be able to detect the correlation between replicas. We explain that neither of these are possible in section \ref{sec:BUs}. The crucial property  is that the Hilbert spaces $\hilb_k^{(n)}$ splits as a direct sum of superselection sectors within which the inner product (and hence the state of radiation) factorises and the swap entropy is equal to the von Neumann entropy, and operations by distant observers can never mix sectors. See \cite{Marolf:2021ghr} for an explanation of the same ideas.

\section{Semiclassical physics in the interior} \label{sec:smooth}

We have verified above that measurements on the radiation give the results expected from a Page curve, explainable by a black hole whose interior degrees of freedom are enumerated by the Bekenstein-Hawking entropy. But to demonstrate that the model evades the information problem (and the small corrections theorem), we must also show that observations including the black hole interior are in line with normal semiclassical expectations. In particular, we would like to check that there is a `smooth horizon', defined in our bit model by verifying the particular entangled state $ \psi_{bc}$ between a bit of radiation $b$ and its interior partner $c$.

At first sight this seems straightforward to check in our model, since our states are still described by the same degrees of freedom as perturbative gravity. However, modifying the inner product introduces some subtleties when we want to describe measurements which include the interior states labelled by $\mathbf{c}$. In this section we explain what these subtleties are, though we will ultimately not attempt to completely understand the nature of interior operators here. Instead, for this paper we have a more modest aim, giving one simple and straightforward definition of interior operators, and showing that it gives results with exponentially small deviation from the model without replica wormholes. This will apply when the number of replicas $n$ is not exponentially large. In particular, this demonstrates that we can simultaneously measure the state at the horizon and a permutation operator on the radiation, and find predictions in line with a smooth horizon and a Page curve.


\subsection{Operators on the physical Hilbert space}

For our purposes, ordinary semiclassical physics (obtained by perturbation theory around the black hole background) corresponds to states and observables in the original `single replica' Hilbert space $\hilb_k$ of section \ref{sec:singleBH}. We would like to compare expectation values of Hermitian operators $\op$ acting on $\hilb_k^{\otimes n}$ to the corresponding expectation values in  our physical Hilbert space $\hilb_k^{(n)}$, to check whether our modified inner product has  interfered with this ordinary physics. Since the underlying vector space of our physical Hilbert space $\hilb_k^{(n)}$ is the same as that for $\hilb_k^{\otimes n}$, one might think that we can simply use the same operator $\op$ as the corresponding observable on $\hilb_k^{(n)}$.

This doesn't quite work for two reasons. 
 First, $\op$ is no longer Hermitian with the modified inner product on $\hilb_k^{(n)}$. Secondly, if the physical inner product $\eta$ has a nontrivial kernel, $\op$ may not even be well defined on $\hilb_k^{(n)}$ (since it will not typically map null states to null states). More abstractly, as a vector space the physical Hilbert space is the quotient
\begin{equation}
	\hilb^{(n)}_k \simeq \frac{\hilb_k^{\otimes n}}{\ker \eta},
\end{equation}
so states are cosets which differ by vectors of zero norm.

So, we would like a better way to construct a well-defined Hermitian operator $\tilde{\op}$ on $\hilb_k^{(n)}$ from a Hermitian operator $\op$ on $\hilb_k^{\otimes n}$. A simple way to do this is to choose a linear map $V:\hilb_k^{\otimes n} \to \hilb_k^{(n)}$, and then define
\begin{equation}\label{eq:Otilde}
	\tilde{\op} = V \op V^\dag \,.
\end{equation}
The resulting $\tilde{\op}$ is an operator on $\hilb_k^{(n)}$, and is manifestly Hermitian (with respect to the inner product $\llangle \cdot|\cdot\rrangle$) if $\op$ is Hermitian (with respect to $\langle \cdot|\cdot\rangle$).

The most simple and obvious candidate for the map $V$ acts trivially on the defining basis states (as $V|\mathbf{c}_1;,\ldots;\mathbf{c}_n\rangle = |\mathbf{c}_1;,\ldots;\mathbf{c}_n\rrangle \,$), but we'll consider another possibility which turns out to have slightly nicer properties (the drawbacks with the naive version are explained in appendix \ref{app:badrecon}). It's easiest to define this through its adjoint $V^\dag$, which we take to be the square root of the inner product matrix $\eta$:
\begin{equation}
	\begin{gathered}
		V^\dag:\hilb_k^{(n)}\to \hilb_k^{\otimes n}, \\
		V^\dag|\mathbf{c}_1;,\ldots;\mathbf{c}_n\rrangle = \eta^\frac{1}{2}|\mathbf{c}_1;,\ldots;\mathbf{c}_n\rangle \,.
	\end{gathered}
\end{equation}
Here $\eta^\frac{1}{2}$ is the unique positive semi-definite Hermitian matrix squaring to $\eta$ (using positive semi-definiteness of $\eta$ itself), and $V^\dag$ is well-defined on $\hilb_k^{(n)}$ because $\eta^\frac{1}{2}$ annihilates null states. From this we can compute $V$ itself using the defining relation $\langle \phi|V^\dag|\psi\rrangle = \overline{\llangle \psi|V|\phi\rangle}$, along with the inner product $\llangle \psi|\phi\rrangle = \langle \psi|\eta|\phi\rangle$ and Hermiticity of $\eta$. We find that $V$ is an inverse square root of $\eta$,
\begin{equation}
	\begin{gathered}
		V:\hilb_k^{\otimes n} \to \hilb_k^{(n)}, \\
		V|\mathbf{c}_1;,\ldots;\mathbf{c}_n\rangle = \eta^{-\frac{1}{2}}|\mathbf{c}_1;,\ldots;\mathbf{c}_n\rrangle \,,
	\end{gathered}
\end{equation}
where $\eta^{-\frac{1}{2}}$ is defined in the obvious way on eigenvectors of $\eta$ with positive eigenvalue, and gives zero on null states in $\ker\eta$.\footnote{More precisely, the definition of adjoint tells us that $\eta V = \eta^{\frac{1}{2}}$, where the factor of $\eta$ on the left comes from the physical inner product. This leaves the definition of $V$ ambiguous up to addition of any element of $\ker\eta$, so we should think of $V$ as mapping to a coset in $\frac{\hilb_k^{\otimes n}}{\ker \eta}$; the definition in the main text gives one representative of the coset.}

Combining these, we find that $V^\dag$ is an isometry since
\begin{equation}\label{eq:Vdagiso}
	VV^\dag = \id,
\end{equation}
but $V$ is not isometric since $V^\dag V$ annihilates null states, and is in fact a projection orthogonal to $\ker\eta$. The isometry property \eqref{eq:Vdagiso} is nice, because it means that our map \eqref{eq:Otilde} takes the identity operator on $\hilb_k^{\otimes n}$ to the identity on $\hilb_k^{(n)}$: $\tilde{\id} = V\id V^\dag = \id$. The result is that measurements that do not act on the black hole interior (operators on the radiation, for example) are not changed by our modified inner product. Not all definitions of $V$ (such as taking a different power of $\eta$) will share this property; see \ref{app:badrecon} for an example.

Using all these definitions, we can compute the matrix elements of $\tilde{O}$ in the physical Hilbert space as follows:
\begin{equation}\label{eq:optildeElements}
	\tilde{\op} = V \op V^\dag \implies \llangle\phi|\tilde{\op}|\psi\rrangle = \langle \phi|\eta^\frac{1}{2} \op \eta^\frac{1}{2} |\psi\rangle.
\end{equation}
We will compare this to the `semiclassical' result $\langle \phi| \op |\psi\rangle$.

The above is mathematically very similar to some considerations of `bulk reconstruction', particularly the non-isometric reconstruction of \cite{Akers:2022qdl}, though has a somewhat different interpretation. We comment on the relation in section \ref{ssec:non-iso}.

There may be other well-motivated ways to define semiclassical operators on $\hilb_k^{(n)}$ or maps $V$: our aim here is not to give a definitive unique or preferred construction, only to show existence of a reasonable definition that demonstrates compatibility with semiclassical horizon physics.

\subsection{Bounding deviations}

Our main aim is now to bound the corrections between the formula 
\eqref{eq:optildeElements} for matrix elements of $\tilde{O}$ and the uncorrected operator $\op$. We can do this using the operator norm $\Vert\cdot\Vert$. Specifically, we will show that $\llangle\phi|\tilde{\op}|\psi\rrangle$ is close to $\langle \phi| \op |\psi\rangle$ for all states $|\psi\rangle$, $|\phi\rangle$ if $\Vert \eta-\id\Vert$ is small, and show that this is true for our replica wormhole bit models.

 For normalised states $|\psi\rangle,|\phi\rangle \in \hilb_k^{\otimes n}$, the corrections to matrix elements are bounded by
\begin{equation}
\begin{aligned}
	|\llangle\phi|\tilde{\op}|\psi\rrangle-\langle\phi|\op|\psi\rangle| &\leq \Vert \eta^\frac{1}{2} \op \eta^\frac{1}{2} -\op \Vert \\
	&\leq (2\Vert\eta^\frac{1}{2}-\id\Vert + \Vert\eta^\frac{1}{2}-\id\Vert^2)  \Vert \op\Vert \\
	&\lesssim \Vert \eta -\id \Vert \, \Vert \op\Vert\,.
\end{aligned}
\end{equation}
In the second line we used the triangle inequality $\Vert A+B\Vert \leq \Vert A\Vert+ \Vert B\Vert$ and sub-multiplicativity $\Vert A B\Vert \leq \Vert A\Vert \Vert B\Vert$ properties. In the final line, we have assumed that $\eta$ is is close to the identity in this norm, neglecting quadratic terms. The upshot is that corrections are small if
\begin{equation}
	\Vert \eta -\id \Vert \ll 1,
\end{equation}
and that this quantity bounds the size of the corrections. Moreover, the linear bound in  $\Vert \eta -\id \Vert$ is sharp by choosing $|\phi\rangle$ and $|\psi\rangle$ to be the eigenstate of $\eta$ with eigenvalue furthest from unity, and $\op$ the operator that projects onto this state. This is the same as demanding small corrections to the inner product for all states (so in particular, the same result guarantees that the physical states $|\psi\rrangle,|\phi\rrangle\in \hilb_k^{(n)}$ are exponentially close to normalised).

Note that this bound also applies for states and  operators acting not only on $\hilb_k^{\otimes n}$, but also jointly on another system (especially on some part of the radiation), with $\eta$ acting as the identity on the additional factor.

Now, since $\eta$ is the identity plus contributions from replica wormholes, we would like to bound the operators associated with the wormholes.  From \eqref{eq:constraintisland}, the contribution to $\eta$ from a given island  $\island$ always takes the same form, and  it is straightforward to compute the norm:
\begin{equation}
	\left\Vert \sum_{\substack{\pi\in\Sym(n)\\ \pi\neq \id}} \lambda_\island^{|\pi|} \, \hat{\pi}(\island) \right\Vert = \sum_{\substack{\pi\in\Sym(n)\\ \pi\neq \id}}  e^{-S_\mathrm{gen}(\island) |\pi|},
\end{equation}
where $e^{-S_\mathrm{gen}(\island)} =\lambda_\island 2^{-(k_\island-k)}$ if $k_\island \geq k$, or just $\lambda_\island$ if $k_\island \leq k$. The terms on the right hand side are the coefficients of permutation operators acting on some subset of bits, and these operators have unit norm (since they are unitary). So the triangle inequality guarantees that the sum of the coefficients bounds the norm of the sum. But this is in fact an equality, since a completely symmetric state is a unit eigenvector of every permutation. 

The resulting sum over permutations has a simple exact form,\footnote{%
We compute the sum of $q^{-|\pi|}$ over $\pi\in\Sym(n)$ by noting its close relation to the trace of a projection $P_{\Sym}$ onto the completely symmetric subspace of $\hilb_q^{\otimes n}$, where $\hilb_q$ is a $q$-dimensional Hilbert space. The trace of $P_{\Sym}$ equals the dimension of this subspace,
\begin{equation}
	\Tr P_\mathrm{\Sym} = \dim \Sym^n \hilb_q= \binom{q+n-1}{n}.
\end{equation}
But we can also write $P_\mathrm{\Sym}$ as an average over permutations $\pi$, and take the trace $\Tr\pi =q^{n-|\pi|}$ term by term:
\begin{equation}\label{eq:TrPSymsum}
	P_\mathrm{\Sym} = \frac{1}{n!} \sum_{\pi\in\Sym(n)}\pi \implies \Tr P_\mathrm{\Sym} =\frac{1}{n!} \sum_{\pi\in\Sym(n)} q^{n-|\pi|}.
\end{equation}
Comparing these gives us
\begin{equation}
	\sum_{\pi\in\Sym(n)} q^{-|\pi|} = \frac{q(q+1)\cdots (q+n-1)}{q^n}.
\end{equation}
Our proof works only for positive integers $q$, but since the sum in \eqref{eq:TrPSymsum} is manifestly polynomial in $q$ we are guaranteed the same result for all $q$. } %
 but for us it is enough to note that it is exponentially small in $S_\mathrm{gen}(\island)$, dominated by the $\frac{n(n-1)}{2}$ transpositions (permutations with $|\pi|=1$). In the semiclassical regime, $S_\mathrm{gen}(\island)$ is large for all possible islands (with minimum given by the Bekenstein-Hawking entropy $S$). We may use the triangle inequality to sum over all possible islands, and since there are only polynomially many terms this does not alter the conclusion that $\eta$ is exponentially close to $\id$:
\begin{equation}
	\Vert \eta -\id \Vert \lesssim n(n-1) e^{-S}.
\end{equation}

In summary, expectation values of operators receive only exponentially small corrections relative to their semiclassical results. For example, this includes the operator that projects onto the `smooth horizon' entangled state $\psi_{bc}$ of the last Hawking quantum and its interior partner.

We make two comments here. First, in some cases there are interesting and robust effects arising from such exponentially small corrections. Indeed, the Page curve studied in section \ref{sec:Page} is an example: a large correction to the (swap) entropy manifests by the expectation value of a swap operator (of unit norm) changing from one exponentially small value to another. Second, our bound on corrections applies in the semiclassical limit $S\gg 1$ at fixed replica number $n$. If $n$ becomes exponentially large (specifically, of order $e^{-S/2}$) then operators which act jointly on all replicas may have large corrections. Not coincidentally, $n\sim e^{-S/2}$ is also the number of repetitions of the swap test required to obtain sufficient statistics to distinguish between the Page curve and a larger entropy.

%
%

\section{Superselection sectors, classical statistics and baby universes}\label{sec:BUs}

The unusual nature of the replica wormhole model inner product --- mixing the interiors of widely separated black holes --- naturally raises various questions. Is such apparent non-locality viable in a sensible physical theory? And how  should we interpret the predictions of a model where observations on different  instances of an object are correlated?

To resolve these questions, we make use of a special property of our inner product on $\hilb_k^{(n)}$, or equivalently the operator $\eta_{k,n}$ on $\hilb_k^{\otimes n}$ that defines it. Namely, it decomposes as a convex combination of factorising inner products $\eta_k(\alpha)$ on $\hilb_k$. This means that
\begin{equation}\label{eq:etaalpha}
	\eta_{k,n} = \int d\mu(\alpha)\, \eta_k(\alpha)^{\otimes n},
\end{equation}
where $\eta_k(\alpha)$ is a family of positive semi-definite Hermitian operators on $\hilb_k$ depending on some parameters labelled by $\alpha$, and $d\mu(\alpha)$ is a probability measure on the space of these parameters. In particular, we are integrating over $\eta_k(\alpha)$ with a non-negative weight. Note that we use the same measure and factors $\eta_k(\alpha)$ for every value of $n$. For $n=0$, we have $\eta_{k,0} = \int d\mu(\alpha) =1$, the identity on the trivial one-dimensional Hilbert space.

More directly in terms of the physical inner product, we have
\begin{equation}\label{eq:IPalpha}
	\begin{gathered}
		\llangle \mathbf{c}_1';\cdots;\mathbf{c}_n'|\mathbf{c}_1;\cdots;\mathbf{c}_n\rrangle = \int d\mu(\alpha)\, \llangle \mathbf{c}_1'|\mathbf{c}_1\rrangle_\alpha \cdots\llangle \mathbf{c}_n'|\mathbf{c}_n\rrangle_\alpha,\\
		\text{where }\llangle \mathbf{c}'|\mathbf{c}\rrangle_\alpha = \langle \mathbf{c}'|\eta_k(\alpha)|\mathbf{c}\rangle.
	\end{gathered}
\end{equation}
This defines a family of Hilbert spaces $\hilb_k^\alpha$ with the same basis of states (labelled by $\mathbf{c}$) of the naive Hilbert space $\hilb_k$, but a modified inner product $\llangle \mathbf{c}'|\mathbf{c}\rrangle_\alpha$.

This section will discuss various aspects of this decomposition of the inner product. First, we explore its consequences for measurements on the radiation. In particular, we explain why \eqref{eq:etaalpha} ensures that the correlations between radiation do not lead to an observable violation of locality, and that all measurements on the radiation (not just the swap tests of section \ref{sec:Page}) must be compatible with a Page curve. Then we explicitly construct this decomposition for replica wormhole bit models. Finally, we explain why a consistent quantum theory guarantees \eqref{eq:etaalpha}, by interpreting the positive measure $d\mu(\alpha)$ as arising from a positive semi-definite inner product in a `baby universe Hilbert space'. Incorporating baby universe leads to a larger Hilbert space, in which $\alpha$ labels superselection sectors for the algebra of observables on the radiation.

\subsection{Measurements on the radiation are superselected}

The outcomes of all possible measurements are described by the density matrix. So we use the given form \eqref{eq:IPalpha} of the inner product in the expression \eqref{eq:rhoRkn} for the $n$-replica density matrix of radiation, and obtain\footnote{\label{foot:norms}There is a small subtlety here, since this does not immediately give normalised $\rho_{R_k}(\alpha)$ with unit trace. The trace can be absorbed into the measure, but this comes at the cost of introducing mild $n$-dependence (mild since the fluctuations of the trace are exponentially small). A better resolution is include a normalising factor for each $\alpha$, which can be incorporated in the path integral using a normalising boundary condition. See section \ref{ssec:norms} for further discussion.}
\begin{equation}\label{eq:rhoalpha}
	\begin{gathered}
	\rho_{R_k}^{(n)} = \int d\mu(\alpha)\, (\rho_{R_k}(\alpha))^{\otimes n}, \\
	\text{where }\langle\mathbf{b}|\rho_{R_k}(\alpha)|\mathbf{b}'\rangle = \sum_{\mathbf{c}',\mathbf{c}} \psi_{\mathbf{b}\mathbf{c}}\bar{\psi}_{\mathbf{b}'\mathbf{c}'} \llangle\mathbf{c}'|\mathbf{c}\rrangle_\alpha.
	\end{gathered}
\end{equation}
A density matrix of this form has a simple interpretation: it describes $n$ identical copies of a state $\rho_{R_k}(\alpha)$ where $\alpha$ is a classically random parameter  selected from the probability measure $d\mu(\alpha)$. For this  it is vital that \eqref{eq:rhoalpha} is a convex combination of density matrices (i.e., that each $\rho_k(\alpha)$ is positive semi-definite Hermitian and the measure $d\mu(\alpha)$ is positive).

A decomposition \eqref{eq:rhoalpha} may be called a `de Finetti form' for $\rho_{R_k}^{(n)}$, due to its relevance in the quantum de Finetti theorem. Roughly speaking, this states that a family of $n$-copy density matrices $\rho_{R_k}^{(n)}$ can be written in the form \eqref{eq:rhoalpha} if and only if the outcomes of experiments on any number of copies are invariant under permutations of the copies; see \cite{caves2002unknown} for a precise statement. The relevance of this theorem in the context of black hole information was pointed out and discussed in \cite{Renner:2021qbe}. 

Operationally, the form \eqref{eq:rhoalpha} means that any sequence of measurements will be consistent with a factorised  state of Hawking radiation, the $n$-fold tensor product of single state $\rho_{R_k}(\alpha)$. In particular, the correlations in $\rho_{R_k}^{(n)}$ do not give rise to observable violations of cluster decomposition. The only effect is that the theory does not give a single definite prediction for $\alpha$, but instead a classical probabilistic prediction. We emphasise that repeated experiments are always correlated, which has the probabilistic interpretation that we only ever select a single sample of $\alpha$ from the distribution $d\mu$: it is not possible to make independent samples, and hence it is impossible to observe the classical uncertainty.

In particular, we can interpret the entropies we computed in section \ref{sec:Page}. The operationally defined `swap' entropy comes from the expectation value of a permutation operator $\hat{\tau}_n$, so we have
\begin{equation}\begin{aligned}
	e^{-(n-1)S^\mathrm{swap}_n(R_k)} &= \Tr(\rho_{R_k}^{(n)} \hat{\tau}_n) \\ 
	&=  \int d\mu(\alpha)\, \Tr(\rho_{R_k}(\alpha)^n)  \\
	&= \int d\mu(\alpha) e^{-(n-1)S_n(\rho_{R_k}(\alpha))}.
	\end{aligned}
\end{equation}
In particular, if we take $n$ close to $1$ we have the simple result
\begin{equation}\begin{aligned}
	S^\mathrm{swap}(R_k) &= \int d\mu(\alpha) S(\rho_{R_k}(\alpha)) \\
	&\neq S( \textstyle\int d\mu(\alpha)\rho_{R_k}(\alpha)).
	\end{aligned}
\end{equation}
So the swap entropy (obtained from measurements on the radiation) is the average entropy of the states $\rho_{R_k}(\alpha)$ with the measure $d\mu$. This is contrasted with the second line: the von Neumann entropy of the average state (the Hawking state $\rho^{(1)}_{R_k}$) does not follow the Page curve, but this feature is unobservable \cite{Marolf:2020xie,Bousso:2020kmy,Marolf:2021ghr}.

Furthermore, this average value for the entropy is in fact typical for a random $\alpha$ selected using the measure $d\mu$: the fluctuations are small. To see this, one can compute the variance of $\hat{\tau}_n$ by computing the expectation value of two copies of that operator, acting on $2n$ black holes. The variance comes from the `connected' piece, which means that the two sets of $n$ black holes must by joined by some replica wormhole: the `disconnected' piece computes the square of the average above. But in the semiclassical regime, such replica wormholes are suppressed (relative to the disconnected result) by an exponentially large factor, with no compensating factor coming from a sum over many states.

This also tells us that the swap entropy is physically relevant not only for swap operators, but for any measurement we like on the radiation, even complete state tomography to determine the density matrix. Outcomes of all measurements will be compatible with a single sample radiation density matrix $\rho_{R_k}(\alpha)$, and a typical sample will have entropies following the Page curve. We conclude that the swap entropy of section \ref{sec:Page} is the physically relevant entropy as claimed.

\subsection{Factorising the replica wormhole bit model}\label{sec:factorising}

We now describe this decomposition explicitly  for our bit model. This is easiest to describe by beginning with the final slice $k=K$; more general $k$ is simply obtained by tracing out the last $2^{K-k}$ bits as described later.

As motivation, note that on the time slice $k=K$ the black hole has completely evaporated so its Bekenstein-Hawking entropy should be zero: it should have a unique internal state since there is simply no black hole left. Equivalently, we expect measurements on the final state of all the Hawking radiation to be consistent with a pure state. We can achieve this if the inner products $\llangle \cdot|\cdot \rrangle_\alpha$ all have rank one: that is, they are projections onto a single (unnormalised) state $|\Psi(\alpha)\rangle \in \hilb_K$:
\begin{equation}\label{eq:etaKalpha}
\begin{gathered}
	\eta_K(\alpha) = |\Psi(\alpha)\rangle \langle \Psi(\alpha)|, \quad
	\text{where } \langle \Psi(\alpha)|\mathbf{c}\rangle = \alpha_\mathbf{c}\in\CC,
\end{gathered}	
\end{equation}
which is manifestly positive and Hermitian as required.
The `$\alpha$ parameters' $\alpha_\mathbf{c}$ are the $2^K$ complex numbers giving the wavefunction of $|\Psi(\alpha)\rangle$. The corresponding final state of radiation is the pure state $ \sum_{\mathbf{b}\mathbf{c}} \psi_{\mathbf{b}\mathbf{c}} \alpha_{\mathbf{c}} |\mathbf{b}\rangle$.

 We would like to write $\eta_{K,n}$ as a convex combination of $n$ copies of such projectors, so
\begin{equation}\label{eq:IPalphaint}
	\langle\mathbf{c}_1',\ldots,\mathbf{c}_n' |\eta_{K,n}|\mathbf{c}_1,\ldots,\mathbf{c}_n\rangle  = \int d\mu(\alpha,\bar{\alpha})  \alpha_{\mathbf{c}_1}\cdots \alpha_{\mathbf{c}_n}\bar{\alpha}_{\mathbf{c}'_1}\cdots \bar{\alpha}_{\mathbf{c}'_n} .
\end{equation}
For the Polchinski-Strominger model \eqref{eq:PSIP}, this is very simple. The matrix elements on the left are given by  delta-functions identifying the labels $\mathbf{c}_r$ with some permutation of the labels $\mathbf{c}_{\pi(r)}'$, and summing over all permutations. To achieve this we can integrate integrate over a Gaussian measure
\begin{equation}\label{eq:PSmu}
	d\mu(\alpha,\bar{\alpha}) = \prod_{\mathbf{c}} d\alpha_{\mathbf{c}}d\bar{\alpha}_{\mathbf{c}}  \frac{e^{-|\alpha_\mathbf{c}|^2}}{\pi},
\end{equation}
which gives a sum over all possible Wick contractions between $\alpha_{\mathbf{c}_r}$ and $\bar{\alpha}_{\mathbf{c}_{\pi(r)}'}$. That is, we take a measure such that $|\Psi(\alpha)\rangle$ is a  Gaussian random state  in $\hilb_K$. The typical norm $\langle\Psi(\alpha)|\Psi(\alpha)\rangle$ is the dimension $2^K$ (with fluctuations of order $2^{K/2}$).

To see how we include replica wormholes, we first split our index $\mathbf{c}$ into two sets of labels, writing $\alpha_{\mathbf{c}\mathbf{\tilde{c}}}$, using $\mathbf{c}$ for the `early' interior (on the island) and $\tilde{\mathbf{c}}$  for the remaining `late' interior. With this notation,
 consider a term in $d\mu$ given by multiplying the Gaussian PS measure \eqref{eq:PSmu} by a polynomial in the $\alpha$, $\bar{\alpha}$ parameters:
\begin{equation}\label{eq:dmuRW}
	d\mu(\alpha,\bar{\alpha})  \supset \left(\prod_{\mathbf{c},\tilde{\mathbf{c}}} d\alpha_{\mathbf{c}\tilde{\mathbf{c}}}d\bar{\alpha}_{\mathbf{c}\tilde{\mathbf{c}}}  \frac{e^{-|\alpha_{\mathbf{c}\tilde{\mathbf{c}}}|^2}}{\pi}\right) \frac{\lambda_\island^{|\pi|}}{n!} \sum_{\mathbf{c}_r,\tilde{\mathbf{c}}_r} \alpha_{\mathbf{c}_{\pi(1)}\tilde{\mathbf{c}}_1} \cdots \alpha_{\mathbf{c}_{\pi(n)}\tilde{\mathbf{c}}_n} \bar{\alpha}_{\mathbf{c}_1\tilde{\mathbf{c}}_1} \cdots \bar{\alpha}_{\mathbf{c}_n\tilde{\mathbf{c}}_n}.
\end{equation}
When we evaluate the inner product \eqref{eq:IPalphaint}, contractions with such a contribution to the  measure give us terms
\begin{equation}
	\langle\mathbf{c}_1',\ldots,\mathbf{c}_n' |\eta_{K,n}|\mathbf{c}_1,\ldots,\mathbf{c}_n\rangle \supset \lambda_\island^{|\pi|}\delta_{\mathbf{c}_{\pi(1)}\mathbf{c}_1'}\cdots \delta_{\mathbf{c}_{\pi(n)}\mathbf{c}_n'} \delta_{\tilde{\mathbf{c}}_1\tilde{\mathbf{c}}_1'}\cdots \delta_{\tilde{\mathbf{c}}_n\tilde{\mathbf{c}}_n'}
\end{equation}
($n!$ equal terms cancel the prefactor). These are precisely the matrix elements of island permutation operators $\hat{\pi}(\island)$. There are additional contractions which permute both the island ($\mathbf{c}_r$ indices) and complement ($\tilde{\mathbf{c}}_r$ indices) which we should perhaps include in our definition of the model, though these are unimportant for the quantities we consider here.

From this basic idea, it is clear that a variety of replica wormhole models can be obtained by adding many such terms to alter the measure $d\mu$, while leaving the space of $\alpha$ parameters and corresponding inner products $\eta_K(\alpha)$ unchanged. See \cite{Marolf:2020rpm} for more discussion of these ideas.

The factorised inner products $\eta_k(\alpha)$ at earlier times ($k<K$) are constructed from $\eta_K(\alpha)$ in the same manner as previously: we evolve to the final slice before tracing out the late radiation. For a general model of the Hawking radiation state, this gives us $\eta_k(\alpha) = \Tr_{K-k} (\eta_K(\alpha)\tilde{\rho})$, where $\tilde{\rho}$ is the state of the  $K-k$ late Hawking interior partner modes, and $\Tr_{K-k}$ is the partial trace over these modes. For our simple model of maximally entangled Hawking radiation, we obtain these inner products simply from a partial trace:
\begin{equation}\label{eq:etakalpha}
\begin{gathered}
	\eta_k(\alpha) = \frac{1}{2^{K-k}}\Tr_{K-k} |\Psi(\alpha)\rangle\langle \Psi(\alpha)|,  \\
	\text{or}\quad \langle \mathbf{c}'|\eta_k(\alpha)|\mathbf{c}\rangle =\frac{1}{2^{K-k}} \sum_{\tilde{\mathbf{c}}} \alpha_\mathbf{c \tilde{c}} \bar{\alpha}_{\mathbf{c}' \tilde{\mathbf{c}}},
	\end{gathered}
\end{equation}
where  $\mathbf{c}$ denotes the first $k$ bits and $\tilde{\mathbf{c}}$ the remaining $K-k$.

For the simplest PS models with the measure \eqref{eq:PSmu} where each $\alpha_\mathbf{c \tilde{c}}$ is an independent complex Gaussian, the measure for $\eta_k(\alpha)$ is a complex Wishart distribution with $2^{K-k}$ degrees of freedom: the second line of \eqref{eq:etakalpha} for complex normal $\alpha$ is the definition of this distribution.

Precisely this distribution appeared for the inner products of end-of-the-world (EOW) brane states in the simple topological model in \cite{Marolf:2020xie}. This is not coincidental, since the path integral computing the inner products is  determined by essentially the same sum over geometries. The difference is the  source of the suppression factor $2^{-(K-k)|\pi|}$ for permutations, coming from tracing over late time Hawking radiation for the PS model and from a geometrical action from the topology in the EOW brane model.

Similarly, the inner products of the EOW brane state in the JT gravity `West coast' model \cite{Penington:2019kki} follow the same distribution (after fixing the spectrum of the Hamiltonian), with the generalisation to a more general state $\tilde{\rho}$ of interior partners. In their equation (D.1), the EOW brane labels $i$ are analogous to our early interior partner labels $\mathbf{c}$ while the energy basis labels $a$ are analogous to our late partners $\tilde{\mathbf{c}}$. Their independent complex Gaussians $C_{i,a}$ are then interpreted as the $\alpha_{\mathbf{c}\tilde{\mathbf{c}}}$ parameters. The extra factors coming from $F(H)$ in (D.10) are analogous to our `late interior partner density matrix' $\tilde{\rho}_{\tilde{\mathbf{c}}'\tilde{\mathbf{c}}}$.

Finally, the non-dynamical model of \cite{Akers:2022qdl} takes the same form, except that the state $|\Psi(\alpha)\rangle$ is taken to be normalised so it is Haar random rather than Gaussian (though they emphasise that they interpret the model with some single fixed choice of state). Comparing with our PS model gives the full ensemble of their models a geometric interpretation (we comment on the geometric realisation of Haar instead of Gaussian ensembles in \ref{ssec:norms}).

\subsection{Baby universes}

From what we have said so far, it may seem surprising and mysterious that a decomposition  \eqref{eq:etaalpha} of $\eta_{k,n}$ into factorised inner products with positive measure exists. Permutation invariance of $\eta_{k,n}$ (i.e.~the fact that it commutes with all permutations) is certainly not sufficient to guarantee this. Here we recall  that there is a principled reason why this must be so: namely, that the integral $\int d\mu(\alpha)$ has a Hilbert space interpretation, so a unitary inner product guarantees a positive measure.

The Hilbert space that gives us an interpretation of the measure is the `baby universe' Hilbert space of \cite{Marolf:2020xie,Marolf:2020rpm}, following ideas of \cite{Coleman:1988cy,Giddings:1988cx}. In the current context, this space $\hilb_{\mathrm{BU}}$ describes the state of any number of post-evaporation black hole interiors. We think of these as acting like closed universes left behind from any black holes that have previously evaporated, with the interior splitting off from the exterior  so it is no longer associated with any particular region in the ambient spacetime (as in figure \ref{fig:finalSlice}). Since our inner product mixes these interiors, the state of such baby universes can affect the radiation that emerges from any new black hole.
In our models, $\hilb_K^{(n)}$ describes the state of $n$ interiors so $\hilb_{\mathrm{BU}}$ is essentially the direct sum of these for all possible values of $n$:
\begin{equation}\label{eq:hbu}
	\hbu \supset \bigoplus_{n=0}^\infty \hilb_K^{(n)} \,.
\end{equation}
(We have written $\supset$ because it's convenient to also include states in $\hbu$ with time-reversed `anti' BUs, described by the dual space $\hilb_K^{(\bar{n})*}$. These won't play a major role for us.) A distinguished element of $\hbu$ is the state  $|\HH\rrangle$ without any baby universes (i.e., a state in the one-dimensional $n=0$ sector of \eqref{eq:hbu}). 

For the simplest case of the PS model, each $n$-universe Hilbert space is the symmetric product $\hilb_K^{(n)}=\Sym^n \hilb_K$ of a single universe Hilbert space, so the full $\hbu$ is simply the bosonic Fock space built on $\hilb_K$ ($\oplus \hilb_K^*$ to include anti-BUs, giving a Hilbert space of $2^K$ complex bosons).

To explain the relevance of these, consider what happens when we create additional black holes in the presence of BUs. To do this we augment our Hilbert spaces $\hilb_k^{(n)}$ of $n$ black holes on a Cauchy surface $\Sigma_k$ to a space $\hat{\hilb}_k^{(n)}$, which additionally allows any number of baby universes.\footnote{In the notation of \cite{Marolf:2020xie} this is analogous to a Hilbert space with asymptotic boundary $\hilb_\Sigma$ where $\Sigma$ has $n$ connected components, for example the space denoted $\hilb_{0,n}$ in the topological model.} This means that states of $\hat{\hilb}_k^{(n)}$ are superpositions of $|\Phi;\mathbf{c}_1;\cdots;\mathbf{c}_n\rrangle $, where $|\Phi\rrangle\in\hbu$ and each $\mathbf{c}_r$ is a $k$-tuple of bits as usual. The previous Hilbert space $\hilb_k^{(n)}$ can be identified with the states $|\HH;\mathbf{c}_1;\cdots;\mathbf{c}_n\rrangle $ without baby universes. But more generally, the inner product on $\hat{\hilb}_k^{(n)}$ mixes the $n$ black holes with the  state $|\Phi\rrangle$ of closed universes, so $\hat{\hilb}_k^{(n)}$ is not simply the tensor product of $\hilb_k^{(n)}$ with $\hbu$. We compute the inner product $\llangle \Phi';\mathbf{c}_1';\cdots;\mathbf{c}_n'|\Phi;\mathbf{c}_1;\cdots;\mathbf{c}_n\rrangle$ as before by evolving the $n$ black holes to the final slice, tracing out the resulting radiation, and then using the inner product on $\hbu$ with a state now containing $n$ extra BUs. In the $m$-universe sector of $|\Phi\rrangle$, $|\Phi'\rrangle$ this will include  islands which mix all $m+n$ black holes.

To make sense of this inner product, we can express it using a convenient basis of `alpha states' for $\hbu$. The basis states $|\alpha\rrangle$ are labelled by the same $2^K$ complex numbers $\alpha_\mathbf{c}$ introduced in \eqref{eq:etaKalpha}, and form an orthonormal (continuum normalised) basis:  $\llangle\alpha'|\alpha\rrangle= \delta(\alpha-\alpha')$, where the delta function is appropriate to the flat measure on $\CC^{2^K}$. If we think of $|\mathbf{c}_1,\ldots,\mathbf{c}_n\rrangle$ as the `number basis' of $2^K$ complex harmonic oscillators (where the occupation numbers count the $\mathbf{c}_r$ equal to any particular value), the states $|\alpha\rrangle$ form the `position basis'.

Using this basis, the inner product on $\hat{\hilb}_k^{(n)}$ is given by
\begin{equation}\label{eq:IPalphac}
	\llangle \alpha';\mathbf{c}_1';\cdots;\mathbf{c}_n'|\alpha;\mathbf{c}_1;\cdots;\mathbf{c}_n\rrangle = \delta(\alpha-\alpha') \llangle \mathbf{c}_1'|\mathbf{c}_1\rrangle_\alpha \cdots \llangle \mathbf{c}_n'|\mathbf{c}_n\rrangle_\alpha \,,
\end{equation}
so the matrix of inner products is block diagonal, and each block factorises into a product over the $n$ black holes separately. In other words, the Hilbert space $\hat{\hilb}_k^{(n)}$ is a direct sum (or more precisely, direct integral) over superselection sectors labelled by $\alpha$, and each superselection sector is an n-fold tensor product over $\hilb_k^\alpha$:
\begin{equation}\label{eq:hilbhatkn}
	\hat{\hilb}_k^{(n)} = \bigoplus_\alpha (\hilb_k^\alpha)^{\otimes n}.
\end{equation}
While the inner product on $\hat{\hilb}_k^{(n)}$ does not factorise, this equation is the next best thing: it splits into superselection sectors which do factorise.

This is very close to our expression \eqref{eq:IPalpha} for the inner product on $\hilb_k^{(n)}$: the last thing to explain is the measure $d\mu(\alpha)$. This follows from embedding $\hilb_k^{(n)}$ into the larger Hilbert space $\hat{\hilb}_k^{(n)}$, and then rewriting the inner product in the $\alpha$ basis. To do this we need the wavefunction of the special state $|\HH\rrangle\in\hbu$ in the $\alpha$ basis, which we write as
\begin{equation}\label{eq:HHwavefunction}
	|\HH\rrangle = \int d\alpha \, e^{i\theta(\alpha)}\sqrt{p(\alpha)} |\alpha\rrangle
\end{equation}
with $p(\alpha)>0$ (and $d\alpha$ the flat measure on $\CC^{2^K}$). With this, we have
\begin{equation}
	\begin{aligned}
	\llangle \mathbf{c}_1';\cdots;\mathbf{c}_n'|\mathbf{c}_1;\cdots;\mathbf{c}_n\rrangle	&= \llangle\HH; \mathbf{c}_1';\cdots;\mathbf{c}_n'|\HH;\mathbf{c}_1;\cdots;\mathbf{c}_n\rrangle \\
	&= \int d\alpha d\alpha' \, e^{i(\theta(\alpha)-\theta(\alpha'))}\sqrt{p(\alpha)p(\alpha')} \llangle\alpha'; \mathbf{c}_1';\cdots;\mathbf{c}_n'|\alpha;\mathbf{c}_1;\cdots;\mathbf{c}_n\rrangle \\
	&= \int d\alpha\, p(\alpha)  \llangle \mathbf{c}_1'|\mathbf{c}_1\rrangle_\alpha \cdots \llangle \mathbf{c}_n'|\mathbf{c}_n\rrangle_\alpha
	\end{aligned}
\end{equation}
where in the last line we use \eqref{eq:IPalphac}.

This is precisely our expression \eqref{eq:IPalpha}, with the measure $d\mu(\alpha) = d\alpha \,p(\alpha)$ given by the modulus square of the wavefunction \eqref{eq:HHwavefunction} of $|\HH\rrangle$. But now positivity of this measure is guaranteed by having a consistent unitary quantum mechanics of baby universes! This argument did not use any details of the specific models we study, so is very robust to modifications and generalisations (as long as they retain permutation invariance and unitarity).


\section{Discussion}\label{sec:disc}

\subsection{Canonical interpretation of replica wormholes}

In this paper, the Hilbert space of states of the black hole interior and their inner product have taken centre stage. This contrasts with most previous treatments of replica wormholes, which use the language of the path integral, often focussing on saddle-point geometries with Euclidean or complex signature. As such, our considerations bring us closer to an interpretation of replica wormholes in a  canonical formalism.

We plan to address this canonical language for replica wormholes more fully in future work.  Here, we will highlight one aspect: topology change induces modifications of the inner product by contributing to the constraints arising from gauging of diffeomorphisms.  To see this, note that we can incorporate topology change by adding appropriate `interaction terms' to the Hamiltonian: our island permutation operators $\hat{\pi}(\island)$ can be thought of as an example of such an interaction. If we treat such interactions perturbatively, each term in the perturbative expansion (with some finite number of topology-changing operators inserted) corresponds to a particular topology-changing process. But since diffeomorphisms are gauged, the Hamiltonian (excepting time-evolution at infinity) is part of the constraints which must be imposed to determine the physical space of states and their inner products. In this way, topology change leaves its mark on the physical Hilbert space.

The path integral naturally imposes the constraints in the inner product by summing over geometries related by gauge symmetries. One way to think about this in a canonical framework is the idea of `group averaging' \cite{Ashtekar:1995zh,Marolf:2000iq}, where a gauge-invariant inner product is constructed by averaging over a gauge group. It is tempting to interpret our expression \eqref{eq:IPconstr1}  for the inner product $\eta$ in this way, as a sum over gauge transformations. In fact, in the Polchinski-Strominger model of section \ref{sec:PS}, this is rather straightforward and standard if we interpret the inner product as a sum over  diffeomorphisms of space which permute interiors (interpreted as disconnected closed universes). We therefore suggest that replica wormhole induced inner products can be interpreted as an extension of gauged `diffeomorphisms', now including permutations of parts of space (so the diffs in question are discontinuituous at the boundaries of these parts). We will explore this idea in future work.

\subsection{Null states and state counting}\label{ssec:null}

We have already commented in several places that the inner product $\eta$ induced by topology change might have a non-trivial kernel. This means that there are `null states': superpositions of states with different semi-classical descriptions with zero physical norm. The idea that null states can emerge from non-perturbative effects (and in particular the constraints) was explored in \cite{Jafferis:2017tiu,Marolf:2020xie}.

This aspect was not crucial for this paper, since here we were primarily concerned with a few simple topologies (in particular a small number $n$ or replicas) in a semiclassical regime, so that $\eta$ is close to the identity. But we believe that these null states are crucial in more interesting circumstances: for large numbers of replicas (which in particular is necessary to describe an $\alpha$ state as introduces in section \ref{sec:BUs}) or close to evaporation. 

Indeed, this is the crux of the information problem. Semiclassical gravity appears to have too many states (as defined by perturbative quantum fields on any self-consistent semiclassical background). Either something is wrong with most of those states or the physical inner product on them is degenerate: that is, there are null states.   But even in the simple models here, it is apparent that null states can emerge from geometry once non-perturbative effects are incorporated. Furthermore, the above discussion interpreting these effects as extended gauge symmetries gives us a reason to expect null states and an interpretation for them: they result from different representations of a state related by gauge transformations, as in \cite{Jafferis:2017tiu}.

Once we have identified that there might be null states, an obvious question is whether we can quantify them: how many states are left in the physical Hilbert space? The models introduced here are simple enough that we can do this directly, by computing the rank of $\eta$.

In forthcoming work \cite{stateCounting}, we will carry out this exercise for the simple PS model \eqref{eq:PSIP}. The result is that for very large $n$, the dimension of $\hilb_\mathrm{phys}$ scales as a power $d^n$, so we can interpret $d$  as the number of states associated with any one black hole. Furthermore, we find $d=e^{S}$ where $S$ is the swap entropy we found in section \ref{sec:Page}, following the Page curve; this means that all but an exponentially small fraction of perturbative states are null after the Page time (i.e., for $k>\frac{K}{2}$). But this is not an indirect calculation via a swap test or replica trick (e.g., the Page curve gives only an indirect lower bound on the number of states via the entanglement of a particular state of black hole and radiation). Instead, it is a `microscopic' state counting, in which we can identify precisely which wavefunctions contribute and which are null states. Perhaps more interesting is the way in which the answer is calculated. We find that for large replica number $n$ of order $e^{\frac{S}{2}}$, a collective continuum description emerges for the `gas of baby universes'. 

\subsection{Factorising models of gravity}

In the light of specific holographic models of gravity such as $\mathcal{N}=4$ Yang-Mills, it may be hard to accept the straightforward interpretation of replica wormholes we have followed here due to the non-factorisation of the inner product for separate systems. The apparent failure of factorisation due to spacetime wormholes has been discussed in the context of holography for some time \cite{Maldacena:2004rf,ArkaniHamed:2007js}, and many perspectives have been advocated recently in light of replica wormholes and other results relying on wormholes \cite{Saad:2019lba,Pollack:2020gfa}

We take the point of view that UV complete, microscopic theories of gravity may indeed have factorising inner products, but gravitational low energy effective field theory is insufficient to resolve the precise inner product. Instead, the latter manages a remarkable feat, remaining internally consistent by describing statistics over a space of superselection sectors (an `ensemble' of duals in AdS/CFT language). Simple statistics (i.e., low moments of the ensemble) may be computable from some relatively simple wormhole geometries, with more complicated geometries suppressed by some power of $e^{-S}$ so the sum over geometries yields a sensible asymptotic expansion.

Despite this, there is good reason to take non-factorising models seriously in efforts to understand gravitational physics (such as the description of black hole interiors). This reason is that the `ensemble' (i.e., the measure $d\mu(\alpha)$ over superselection sectors) is not a fixed parameter of the theory, but rather a property of the state; specifically, it is the square of the wavefunction of the state of closed (baby) universes. In section \ref{sec:BUs} we considered only the `Hartle-Hawking' or no-boundary state, which is a natural (and usually implicit) choice, but it is certainly not the only possibility. In particular, we can ask for a state whose wavefunction is more narrowly peaked close to a specific $\alpha$-state, and by taking a narrower wavefunction we can approach closer and closer to a specific superselection sector. This state might be the result of a physical process by which an observer discovers the $\alpha$ parameters through measurements on the radiation from many black holes. Furthermore, these states have geometric descriptions as populations of many baby universes. Given all this, it is reasonable that a geometric description of a single typical $\alpha$-state might be a guide to the physics in microscopic factorising theories.

If we straightforwardly follow this for our models, we would arrive at one of the factorised inner products $\llangle \cdot |\cdot\rrangle_\alpha$ as  in \eqref{eq:IPalpha}. And in \eqref{eq:etaKalpha} we saw that this inner product takes the form of a projection onto a single state of the complete black hole interior after evaporation. As such, the result is very similar to the final state proposal \cite{Horowitz:2003he}. But for us, the projection emerges from computation of an inner product in a standard quantum system, and does not appear to involve any modification of the rules of quantum mechanics. However, a complete description of the theory (and in particular, the correct prescription for predicting the experience of infallers) requires us to properly identify the observables in the interior, as discussed in the next subsection.

If we construct interior observables following the na\"ive approach of section \ref{sec:smooth} using the inner product $\llangle \cdot |\cdot\rrangle_\alpha$ for some fixed $\alpha$, we will in fact find large deviations from semiclassical physics after the Page time (though see comments in section \ref{ssec:subspaceDep}). This is not in contradiction with the results of section \ref{sec:smooth}, because fixing an $\alpha$ state takes us to a regime of large replica number $n$. But this breakdown of semiclassical physics happens for a principled reason (see also the discussion in section 7.2 of \cite{Marolf:2020rpm}). While more complicated topologies are individually suppressed by powers of $e^{-S}$, for large $n$ there is also a proliferation of the number of geometries contributing at any order. Once $n\gtrsim e^{\frac{S}{2}}$, the combinatorics overwhelm the usual suppression, and the topological expansion no longer gives a sensible asymptotic series. That is, if we truncate the series with an upper bound on the counting parameter $|\pi|$ (analogous to the Euler characteristic in a two-dimensional model, for example), the first omitted term is not smaller than the included terms. Crucially, this failure of semiclassical physics does not constitute a paradox, since it is a consequence of concrete non-perturbative gravitational effects invalidating the semiclassical approximation. And these effects do not endanger ordinary physics since they become important only in a region of spacetime which violates an entropy bound, with the $\log$ of the number of possible semiclassical states in the region exceeding the area of its boundary in Planck units.

Having said all this, we do not necessarily conclude that factorising models fail to have a smooth horizon, because we are not confident that the approach of section \ref{sec:smooth} is the physically relevant prescription (the ideas in \ref{ssec:subspaceDep} may lead towards a different conclusion, for example). The point is only that a firewall scenario need not be in contradiction with gravitational effective field theory once non-perturbative effects are included. To make more definitive statements, we require a well-justified model for infalling observers and interior measurements, which we now comment on.

\subsection{The infalling state and observables in the interior}\label{ssec:infallers}

In section \ref{sec:smooth}, we described a rather abstract procedure for constructing interior operators in the physical Hilbert space $\hilb_\mathrm{phys}$. While this is plausible (and provides proof of concept that a smooth horizon is compatible with a Page curve), it has little justification as a realistic model for infalling observers. To rectify this, we require a model which allows for an interesting state of infalling matter, not just taking the infalling vacuum as for the models presented here. In particular, this state could include an observer and their experimental apparatus, whose dynamics can describe some measurement process.

Here we just point out one challenge to building such a model by straightforwardly following the ideas of this paper. Suppose we begin with a perturbative Hilbert space $\hilb_\mathrm{pert}$ including infalling $\mathbf{a}$ modes outside the black hole, and an evolution which freely maps those $\mathbf{a}$ modes to left-moving interior modes without otherwise modifying their state. Define a physical Hilbert space $\hilb_\mathrm{phys}$  with inner product $\eta$ as before, by following semiclassical evolution up to some late time and then summing over replica wormhole operators. Then the infalling $\mathbf{a}$ modes will receive similar corrections to their inner product as the Hawking partner $\mathbf{c}$ modes. But if the evolution is required to be unitary in $\hilb_\mathrm{phys}$, the $\mathbf{a}$ modes must have the same corrections before they enter the black hole. Since these begin as excitations at a great distance (as the in-state at past infinity, perhaps), it seems that we should expect non-perturbative gravitational effects even far from the black hole.

It is surprising at first sight that the inner product in a weakly curved part of spacetime appears to receive non-perturbative gravitational corrections, which furthermore depend on whether the excitations in question eventually fall into a black hole (which may not even have formed yet). In fact, if these corrections do not fall off with distance they may be in conflict with cluster decomposition (and not only in a mild unobservable way accounted for by superselection sectors). This is perhaps worst in AdS, where the formation of a black hole at some later time can depend on whether we turn on sources at the boundary in the future (making unitary deformations of the Hamiltonian in a CFT dual). If the inner product depends on such choices, it constitutes a violation of boundary unitarity.

We therefore expect a unitary model of infallers to either require modifications to the inner product of infalling modes, or a nontrivial evolution taking them from far away into the black hole, or a combination of the two. Whatever is required, we would like a model inspired by concrete gravitational dynamics computable in effective field theory. In that context, the crucial question is whether the unitary evolution of perturbative gravity is preserved by non-perturbative topology changing effects (and if so, how).

\subsection{Interior operators are not superselected}

If we wish to ask about the experience of infalling observers in a model with $\alpha$ states, it is important to appreciate that the argument of superselection sectors does not apply to the relevant operators.

In section \ref{sec:BUs}, we argued that the failure of factorisation can be accounted for by taking a factorised inner product $\llangle\cdot|\cdot\rrangle_\alpha$, but with $\alpha$ selected from some classical probability distribution (giving rise to classical correlations between radiation from different black holes). Equivalently, we can say that (1) the $n$-replica Hilbert space splits as the direct sum of superselection sectors labelled by $\alpha$ (as in \eqref{eq:hilbhatkn}), (2) each such superselection sector factorises as a tensor product over the $n$ replicas, and (3) the operators acting on the radiation are superselected. Part (3) means that these operators do not mix superselection sectors labelled by different $\alpha$, so they are block diagonal matrices with respect to the direct sum decomposition. This follows from the general arguments given in \cite{Marolf:2020xie}.

However, it is crucial that the argument for (3) only applies for asymptotic observables such as operators acting on the radiation. There is no such argument for more general operators such as those which act in the black hole interior directly; these operators might mix different $\alpha$ sectors. As such, we cannot use language of superselection sectors or ensembles when we discuss such operators. In paritcular, the expectation value in the Hartle-Hawking state of baby universes need not be the average of expectation values in each $\alpha$ state.

This should be borne in mind whenever we discuss operators that do not have a purely asymptotic definition in terms of boundary conditions at infinity. For example, it is not obvious that the ensemble language  is applicable to the firewall discussion of \cite{Stanford:2022fdt}.

\subsection{Subspace dependence of interior operators}\label{ssec:subspaceDep}

The construction of operators we used in section \ref{sec:smooth} has a curious feature: it depends on the subspace of states on which we choose to define the physical operator $\tilde{\op}$ corresponding to a given operator $\op$ in the perturbative Hilbert space.

Concretely, suppose that $\op$ acts on the $\hat{k}$th bit in the interior of some black hole. This means that we can regard $\op$ as an operator on the Hilbert space $\hilb_{k}$ of the black hole at time $t_{k}$ for any $k\geq \hat{k}$. Given a choice of $\hilb_{k}$, we can use the physical inner product $\eta_{k,n}$ defining the $n$-replica physical Hilbert space $\hilb_{k}^{(n)}$ to construct $\tilde{\op}$ acting on that space. But the resulting physical operator depends on $k$ in an essential way: for example, the expectation value of $\tilde{\op}$ in the simple Hawking state created by the free unitary evolution $U$ is not independent of $k$. At a technical level, this happens because the matrix elements \eqref{eq:optildeElements} of $\tilde{\op}$ are not linear in the inner product $\eta$. The result is that removing some bits (by placing them in the Hawking state created by $U$ before tracing out) does not commute with the map from $\op$ to $\tilde{\op}$. While this may seem problematic at first sight, we think that this effect is physical.

To motivate this, we suppose we have $n$ black holes and would like to define an operator $\tilde{\op}$ inside a specific one of them, and consider what happens at various times. Taking an extreme case, we can allow the black holes to evaporate completely (taking $k=K$). At that point the black hole interiors have essentially become $n$ closed universes, no longer associated with any particular exterior, and without additional information are indistinguishable (with symmetric wavefunction). It therefore doesn't make sense to assign an operator to any one particular interior (our construction $\tilde{\op}$ might act by averaging over the $n$ possible interiors, for example). On the other hand, at early times (small $k$) when the black holes are young and macroscopic, their interiors and exteriors are geometrically attached. In this case we expect to be able to distinguish the specific interior we are interested in (at least to high accuracy) by means of the (distinguishable) exterior to which it is attached. But the latter case can be obtained from the former by tracing out all of the late Hawking radiation and interior partners: this tracing over a subsystem has had a physical effect on our ability to identify one specific interior.

How are we able to identify one specific interior in the case of the young black hole? Our answer is that we are not assuming a generic state, but rather a very special state that connects a given interior and exterior; in particular this state involves coherent entanglement between interior and exterior of many short-distance modes. At late time these modes are stretched into Hawking radiation and partners, and the late time Hilbert space makes no assumptions about their state. For the young black hole, we may use the large entanglement in these short-distance modes to `dress' the operator $\tilde{\op}$. Even after the black hole has evaporated, we can continue to dress the operator $\tilde{\op}$ by reference to the entangled Hawking state, and hence identify a specific interior. By assuming the Hawking state for most of the radiation and interior partners, we can identify one particular interior as the one which is highly entangled with some specific set of Hawking radiation.

We can also apply these ideas in the case of an $\alpha$-state (and hence in a factorising model). In this case, the matrix elements of a physical operator $\tilde{\op}$ are close to the those of the corresponding $\op$ acting on the perturbative Hilbert space if we define it to act on fewer than half the interior modes (and can thus use the remaining state for `dressing'). Such operators will give find predictions in line with semiclassical expectations. The requirement that we use half the bits for dressing means that this works before the Page time, but there is nothing particularly special about that time: we can use \emph{any} of the interior bits for dressing, as long as we have more than half in total. In conclusion, we can always define interior operators in line with semiclassical expectations as long as we do not attempt to talk about more than half of the interior bits at once.

While this gives a consistent way in which to talk about interior operators, it is not clear whether such operators are relevant for the experience of infalling observers. As discussed in section \ref{ssec:infallers} we require a model of infalling dynamics for such questions.

These ideas invite comparison to quantum error correction \cite{Verlinde:2012cy,Almheiri:2014lwa}, state dependence \cite{Papadodimas:2015jra} or state-specific reconstruction \cite{Akers:2021fut,Akers:2022qdl}, and other discussions of non-perturbative dressing \cite{Jafferis:2017tiu}. It would be interesting to make such connections precise.

\subsection{Relation to non-isometric reconstruction}\label{ssec:non-iso}

Some of the ideas in this paper, in particular those of section \ref{sec:smooth}, bear some similarity to the non-isometric reconstruction discussed in \cite{Akers:2022qdl}. At a technical level, our perturbative and physical Hilbert spaces $\hilb_\mathrm{pert}$ and $\hilb_\mathrm{phys}$ are analogous to the bulk and boundary Hilbert spaces $\hilb_\mathrm{bulk}$ and $\hilb_\mathrm{boundary}$ respectively of \cite{Akers:2022qdl}. This analogy is closest in section \ref{sec:smooth} where we define a linear map $V:\hilb_\mathrm{pert}\to \hilb_\mathrm{phys}$, which invites comparison with the reconstruction map $V:\hilb_\mathrm{bulk}\to \hilb_\mathrm{boundary}$ of \cite{Akers:2022qdl}. But here we will emphasise some important differences in interpretation.

First, we do not invoke any notion of duality. Both of our Hilbert spaces $\hilb_\mathrm{pert}$ and $\hilb_\mathrm{phys}$  are to be understood as `bulk' constructions, with states defined by wavefunctions of gravitational variables. While our $\hilb_\mathrm{pert}$ is rather similar to $\hilb_\mathrm{bulk}$ (containing the states of perturbative bulk effective field theory),  \cite{Akers:2022qdl} do not posit any realisation of $\hilb_\mathrm{boundary}$ using the same variables.

Perhaps most importantly, from our perspective $\hilb_\mathrm{pert}$ does not ultimately have any physical significance. For us it arises from a theory of gravity which explicitly neglects non-perturbative effects. The only physical Hilbert space is $\hilb_\mathrm{phys}$. If some physics of $\hilb_\mathrm{pert}$ (perhaps a smooth horizon, for example) does not survive when we pass to the physical Hilbert space $\hilb_\mathrm{phys}$, that is not a problem: we would be inclined to conclude that perturbative gravity fails due to some concrete non-perturbative gravitational process. In particular, since we do not hold particular attachment to $\hilb_\mathrm{pert}$ we do not see any obvious motivation to map observables $\op$ in $\hilb_\mathrm{pert}$ non-linearly to operators $\tilde{\op}$ in $\hilb_{\mathrm{phys}}$.

Finally, the central object in our discussion was the physical inner product $\eta$ (a sesquilinear form or equivalently a semi-definite Hermitian operator on $\hilb_\mathrm{pert}$). We introduced the map $V:\hilb_\mathrm{pert}\to \hilb_\mathrm{phys}$ only later in order to define a simple construction of physical operators (and as already discussed above, we are not confident that our construction of $V$ is physically well-motivated). In contrast, the analogous $V:\hilb_\mathrm{bulk}\to \hilb_\mathrm{boundary}$ was the central object in \cite{Akers:2022qdl}.\footnote{The obvious analogue of $\eta$ for them is $V^\dag V$, but this is imperfect since for us $V^\dag V\neq \eta$; instead $V^\dag V$ is a projection orthogonal to the kernel of $\eta$.}

\subsection{Normalisation of states}\label{ssec:norms}

We were not very careful about normalising all our states, since this was not very important for our considerations: the norm of the states of interest were always exponentially close to unity, and the variance in the norm of $\alpha$ states in section \ref{sec:BUs} is exponentially small. But for some considerations, it is important to get these details correct.

One might simply normalise all states in the physical inner product. But as remarked in footnote \ref{foot:norms} and in \cite{Marolf:2020rpm}, this leads to a mild $n$-dependence in the measure of $\alpha$ states which does not seem physical. Perhaps a better approach is to use the decomposition \eqref{eq:rhoalpha} into superselection sectors and then to normalise independently in each $\alpha$ state.

This can in fact be achieved without using the full decomposition into $\alpha$ sectors, by using appropriate `normalising' boundary conditions for the path integral. The idea is that for each replica of the radiation density matrix $\rho_R$ in a computation, we also insert a boundary condition representing a factor of $\frac{1}{\Tr\rho_R}$. We can think of this as a `boundary inserting operator' acting on the Hilbert space of baby universes in the language of \cite{Marolf:2020xie}, also discussed in \cite{Marolf:2020rpm} directly in the context of black holes. But it is not immediately obvious how to compute path integrals in the presence of such an operator since it is not polynomial in boundary conditions. We briefly sketch how to do this. First, we write $\Tr\rho_R$ as a disconnected piece $\langle \Tr\rho_R\rangle =1$ (since the Hawking density matrix is normalised) plus a connected piece $(\Tr\rho_R)_c$ which (by definition) must always be joined to some other replica by a wormhole. The wormhole contributions are suppressed so we have $(\Tr\rho_R)_c\ll 1$, which allows us to expand $\frac{1}{\Tr\rho_R} = \frac{1}{1+(\Tr\rho_R)_c}$ as a geometric series. We can truncate this series to get to any desired order in an expansion in $e^{-S}$.

Taking the PS model as an example, these computations of normalised radiation density matrices change the Gaussian $\alpha$ states \eqref{eq:PSmu} to Haar random $\alpha$ states. The normalising denominators produce the subleading Weingartens studied in \cite{Stanford:2021bhl} (where interestingly, a similar geometric series showed up in a different context).

However, at least in the context of AdS/CFT where we can create a black hole state by unitary deformations of the Hamiltonian, we do not expect such normalisation to be necessary. The resolution of the correct recipe is closely tied to the discussion of unitarity for infallers in section \ref{ssec:infallers}.

\subsection{A Lorentzian gravitational path integral?}\label{ssec:LorentzianPI}

We motivated the definition of our inner products by a sum over geometries in Lorentzian signature (with mild topology-changing singularities). This was useful to give us a direct Hilbert space interpretation, as opposed to Euclidean (or complex) metrics which obscure the intermediate states. This accords with an attitude that the gravitational path integral may be best defined over Lorentzian signature geometries; see \cite{Marolf:2022ybi} and references therein. This paper addresses in particular the inclusion of Lorentzian conical singularities relevant for replica wormholes (the `crotch singularity' of \cite{Louko:1995jw}), also discussed in \cite{Colin-Ellerin:2020mva,Colin-Ellerin:2021jev}. These also essential for a recently proposed Lorentzian path integral of JT gravity \cite{Usatyuk:2022afj}. With this perspective, Euclidean or complex metrics appear only as calculational tools, appearing when we choose to deform the contour of integration to pass through such saddle-points. 

However, our models of replica wormholes implement this idea extremely crudely, and miss or ignore important features of realistic theories. Here we point out two related such features. One is that our sum over islands of a given topology (for computing the Page curve in section \ref{sec:QES}, for example) has purely real weightings. More realistically, we expect the integral over the location of the splitting surface in replica wormholes (at finite $n$) to be highly oscillatory from the weighting $e^{iS}$. This means that we expect saddle-points to require deformation to complex metrics (particularly since all equations of motion cannot be obeyed at the splitting surface $\partial \island$ for real Lorentzian metrics). A second feature is that our models only arise if we restrict the allowed islands $\island$ to be bounded by a surface lying near to the event horizon; without such a restriction, islands with $\partial\island$ lying close to the singularity (hence with small area) would be much more important.

We expect that the latter issue is resolved in a more realistic model by the former. That is, if we start with a fluctuating integral over metrics, then the integral over islands moving away from the horizon and close to the singularity will tend to be suppressed by rapid oscillations. We would like to understand the details of this. In particular, assuming this is correct, we can ask whether it applies to all quantities, or only simple exterior observations (like the Page curve)? Perhaps certain calculations (such as complicated exterior observables sensitive to the $\alpha$ state, or bulk interior observables) receive important unsuppressed corrections from physics near the singularity, and hence are outside the remit of low-energy effective gravity.

%
%
%
%
%
%

\subsection*{Acknowledgements}
I would like to thank Samir Mathur for questions and discussions motivating this work. I am also very grateful to Don Marolf for many discussions, comments and encouragement.

I am supported by DOE grant DE-SC0021085 and a Bloch fellowship from Q-FARM. I was also supported by NSF grant PH-1801805, by a DeBenedictis Postdoctoral Fellowship, and by funds from the University of California. This work was performed in part at Aspen Center for Physics, which is supported by National Science Foundation grant PHY-1607611.

\appendix
\addtocontents{toc}{\protect\setcounter{tocdepth}{0}}

\section{Other reconstruction maps}\label{app:badrecon}

In section \ref{sec:smooth}, we made use of a map $V:\hilb_\mathrm{pert}\to\hilb_\mathrm{phys}$, taking states in the perturbative Hilbert space to states in the physical Hilbert space. There is a simple and obvious candidate for this map which acts trivially on the defining basis states:
\begin{equation}
	\begin{gathered}\label{eq:Valt}
		V|\mathbf{c}_1;,\ldots;\mathbf{c}_n\rangle = |\mathbf{c}_1;,\ldots;\mathbf{c}_n\rrangle \,.
	\end{gathered}
\end{equation}
But we chose a different candidate for $V$. Here we briefly explain why we preferred the $V$ defined in the main text.

In the definition \eqref{eq:Valt}, if we are being  precise we should think of the right hand side as an element of the coset vector space $\hilb_\mathrm{phys}\sim \hilb_\mathrm{pert}/\ker\eta$ (so we can always add a null state if they exist). So the resulting $V$ is the quotient map taking any vector to its coset. We can also compute the adjoint $V^\dag$ using the defining relation $\langle \phi|V^\dag|\psi\rrangle = \overline{\llangle \psi|V|\phi\rangle}$, along with the inner product $\llangle \psi|\phi\rrangle = \langle \psi|\eta|\phi\rangle$ and Hermiticity of $\eta$, which gives us
\begin{equation}
	V^\dag |\mathbf{c}_1;,\ldots;\mathbf{c}_n\rrangle = \eta |\mathbf{c}_1;,\ldots;\mathbf{c}_n\rangle.
\end{equation}
This operator is well-defined since adding a null state on the left does not change the result. In particular, we find that $V$ is a square root of the inner product $\eta$ in the sense that
\begin{equation}\label{eq:VVdagalt}
	V^\dag V = \eta.
\end{equation}
We similarly have $VV^\dag = \eta$ (with $\eta$ defining an operator on the coset space $\hilb_\mathrm{phys}$ because it annihilates null states by definition).

However, when we turn to constructing physical operators using \eqref{eq:Otilde} by conjugation of a perturbative operator as $\tilde{\op} = V \op V^\dag $, this result has a strange consequence. Namely, the reconstruction of the identity operator on $\hilb_\mathrm{pert}$ is not the identity on $\hilb_\mathrm{phys}$. Instead, we have $\tilde{\id} = \eta$. The same applies when we allow $\op$ to act on some other Hilbert space (such as the radiation): even if $\op$ acts trivially on the black hole interior, our construction of $\tilde{\op}$ does not leave it invariant. We do not find this physically acceptable: the mere existence of a black hole should not effect the physics of some unrelated, decoupled non-gravitational system elsewhere in the universe.

\bibliographystyle{JHEP}

\bibliography{biblio}

\end{document}